\documentclass[11pt]{article}
\pdfoutput=1
\usepackage[centertags]{amsmath}
\usepackage[square, comma, sort&compress,numbers]{natbib}
\usepackage{array,multirow}
\numberwithin{equation}{section}
\usepackage{amssymb,mathtools}
\usepackage{amsfonts,amsthm,stmaryrd}
\usepackage{graphicx}
\usepackage{color,bm}
\usepackage[normalem]{ulem}

\usepackage{epsfig}
\usepackage{bbold}

\usepackage{wrapfig}
\usepackage{float}
\usepackage{soul}
\usepackage{color}
\usepackage{ulem}

\usepackage{comment}

\usepackage[dvipsnames]{xcolor}

\usepackage{tikz,pgf}
\usetikzlibrary{shapes}
\usetikzlibrary{calc}
\usetikzlibrary{decorations.pathmorphing}
\usetikzlibrary{decorations.pathreplacing,shapes.misc}
\usetikzlibrary{positioning}
\usetikzlibrary{arrows}
\usetikzlibrary{decorations.markings}
\usetikzlibrary{shadings}

\usetikzlibrary{intersections}

\def\sideremark#1{\ifvmode\leavevmode\fi\vadjust{\vbox to0pt{\vss
 \hbox to 0pt{\hskip\hsize\hskip1em
 \vbox{\hsize3cm\tiny\raggedright\pretolerance10000
  \noindent #1\hfill}\hss}\vbox to8pt{\vfil}\vss}}}

\newcommand{\be}{\begin{equation}}
\newcommand{\ee}{\end{equation}}
\newcommand{\ba}{\begin{eqnarray}}
\newcommand{\ea}{\end{eqnarray}}

\def\dps{\displaystyle}

\def\tr{{\rm Tr}}

\advance\textheight\topskip
\renewcommand{\tilde}{\widetilde}

\renewcommand{\simeq}{\cong}

\newcommand{\bref}[1]{\textbf{\ref{#1}}}

\newcommand{\binner}[2]{%
  {\langle}\kern-4.15pt{\langle}#1{,}\,#2{\rangle}\kern-4.15pt{\rangle}}

\newcommand{\half}{\mathchoice{%
    \ffrac{1}{2}}{\frac{1}{2}}{\frac{1}{2}}{\frac{1}{2}}}

\newcommand{\ffrac}[2]{\raisebox{.5pt}%
  {\footnotesize$\displaystyle\frac{#1}{#2}$}\kern1pt}

\numberwithin{equation}{section} \makeatletter
\@addtoreset{equation}{section}

\def\be{\begin{equation}}
\def\ee{\end{equation}}
\def\ba{\begin{array}}
\def\ea{\end{array}}

\def\dps{\displaystyle}
\def\tr{{\rm Tr}}

\newdimen\tableauside\tableauside=1.0ex
\newdimen\tableaurule\tableaurule=0.4pt
\newdimen\tableaustep
\def\phantomhrule#1{\hbox{\vbox to0pt{\hrule height\tableaurule
width#1\vss}}}
\def\phantomvrule#1{\vbox{\hbox to0pt{\vrule width\tableaurule
height#1\hss}}}
\def\sqr{\vbox{%
  \phantomhrule\tableaustep

\hbox{\phantomvrule\tableaustep\kern\tableaustep\phantomvrule\tableaustep}%
  \hbox{\vbox{\phantomhrule\tableauside}\kern-\tableaurule}}}
\def\squares#1{\hbox{\count0=#1\noindent\loop\sqr
  \advance\count0 by-1 \ifnum\count0>0\repeat}}
\def\tableau#1{\vcenter{\offinterlineskip
  \tableaustep=\tableauside\advance\tableaustep by-\tableaurule
  \kern\normallineskip\hbox
    {\kern\normallineskip\vbox
      {\gettableau#1 0 }%
     \kern\normallineskip\kern\tableaurule}%
  \kern\normallineskip\kern\tableaurule}}
\def\gettableau#1 {\ifnum#1=0\let\next=\null\else
  \squares{#1}\let\next=\gettableau\fi\next}

\def\cD{\mathcal{D}}

\def\cG{\mathcal{G}}
\def\cH{\mathcal{H}}

\def\cJ{\mathcal{J}}
\def\cK{\mathcal{K}}
\def\cL{\mathcal{L}}
\def\cM{\mathcal{M}}
\def\cN{\mathcal{N}}
\def\cO{\mathcal{O}}
\def\cP{\mathcal{P}}

\def\cW{\mathcal{W}}

\def\bmb{\bm{b}}

\def\bmd{\bm{d}}
\def\bme{\bm{e}}
\def\bmf{\bm{f}}

\def\bmh{\bm{h}}
\def\bms{\bm{s}}
\def\bmt{\bm{t}}

\def\bml{\bm{l}}
\def\bmm{\bm{m}}
\def\bmn{\bm{n}}

\def\bmI{\bm{I}}

\def\bmT{\bm{T}}
\def\bmU{\bm{U}}
\def\bmZ{\bm{Z}}

\def\bmcK{\bm{\cK}}
\def\bmcJ{\bm{\cJ}}
\def\bmcL{\bm{\cL}}

\def\bmlambda{\bm{\lambda}}

\def\bmzeta{\bm{\zeta}}

\numberwithin{equation}{section} \makeatletter
\@addtoreset{equation}{section}

\newcommand{\bT}{\mathbf{T} }	% su(M) generators
\def\ads{\text{AdS}}

\newcommand{\diffzeta}{\bm \Gamma }

\newcommand{\tnfunc}{\mathbb{TN} }
\newcommand{\scfunc}{\mathbb{SC} }

\def\cft1{\text{CFT}_{1}}

\def\be{\begin{equation}}
\def\ee{\end{equation}}
\def\ba{\begin{array}}
\def\ea{\end{array}}

\def\dps{\displaystyle}

\def\ba{\begin{array}}
\def\ea{\end{array}}

\def\dps{\displaystyle}

\def \rd{{\rm{d}}}

\numberwithin{equation}{section} % Number equations within sections (i.e. 1.1, 1.2, 2.1, 2.2 instead of 1, 2, 3, 4)
\numberwithin{figure}{section} % Number figures within sections (i.e. 1.1, 1.2, 2.1, 2.2 instead of 1, 2, 3, 4)
\numberwithin{table}{section} % Number tables within sections (i.e. 1.1, 1.2, 2.1, 2.2 instead of 1, 2, 3, 4)

\usepackage{sectsty} % Allows customizing section commands
\allsectionsfont{\centering } % Make all sections centered, the default font and small caps

\def\XXint#1#2#3{{\setbox0=\hbox{$#1{#2#3}{\int}$}
     \vcenter{\hbox{$#2#3$}}\kern-.5\wd0}}

\def \Tr{\mbox{Tr\,}}

\def \tr{\mbox{tr\,}}

\usepackage{jheppub}

\makeatletter
\def\@fpheader{\vspace{-.1cm}}
\makeatother

\title{\centering{Schwarzian for  colored Jackiw-Teitelboim   gravity}}

\author[a]{Konstantin\ Alkalaev,} 
\author[b]{Euihun\ Joung,}
\author[c,d,e]{Junggi\ Yoon}

\affiliation[a]{I.E. Tamm Department of Theoretical Physics, 
\\
P.N. Lebedev Physical Institute,
\\ 
Leninsky ave. 53, 119991 Moscow, Russia}
\affiliation[b]{Department of Physics, Kyung Hee University,\\ 26 Kyungheedae-ro Dongdaemun-gu, Seoul 02447, Korea}
\affiliation[c]{Asia Pacific Center for Theoretical Physics,\\77 Cheongam-ro, Nam-gu, Pohang-si, Gyeongsangbuk-do, 37673, Korea}
\affiliation[d]{Department of Physics, POSTECH\\ 77 Cheongam-ro, Nam-gu, Pohang-si, Gyeongsangbuk-do, 37673, Korea}
\affiliation[e]{School of Physics, Korea Institute for Advanced Study\\
85 Hoegiro Dongdaemun-gu, Seoul 02455, Korea}

\emailAdd{alkalaev@lpi.ru}
\emailAdd{euihun.joung@khu.ac.kr}
\emailAdd{junggi.yoon@apctp.org}

\abstract{We study the boundary effective action of the colored version of the Jackiw-Tei\-tel\-bo\-im (JT) gravity. We derive the boundary action, which is the color generalization of the Schwar\-zi\-an action, from the $su(N,N)$ BF formulation of the colored JT gravity. Using different types of the $SU(N,N)$ group decompositions both the zero and finite temperature cases are elaborated. We provide the semi-classical perturbative analysis of the boundary action and discuss the instability of the spin-1 mode and its implication for the quantum chaos. A rainbow-AdS$_2$ geometry is introduced where the color gauge symmetry is spontaneously broken.}

\begin{document}

\maketitle
\flushbottom

\section{Introduction}

In recent years,  Jackiw-Teitelboim~(JT) gravity~\cite{Teitelboim:1983ux,Jackiw:1984je} has enabled us to deepen the understanding of quantum gravity and the black hole information problem. The action of the nearly-AdS$_2$ JT gravity~\cite{Maldacena:2016upp} 
\begin{align}
    S_{JT}\,=\, -\frac{1}{16\pi G_N}\left[\, \int_{\cM_2} \sqrt{|g|}\phi (R+2) + 2\int_{\partial\cM_2 } \sqrt{|h|} \phi(K-1)\,\right]  \label{def: jt action}
\end{align}
reduces to the Schwarzian boundary action which is dual to the low energy effective action of the Sachdev-Ye-Kitaev~(SYK) model~\cite{KitaevTalks,kitaevfirsttalk,Polchinski:2016xgd,Jevicki:2016bwu,Maldacena:2016hyu}. This Schwarzian mode plays a central role in the maximal quantum chaos of the SYK-like models~\cite{KitaevTalks,kitaevfirsttalk,Maldacena:2016hyu,Fu:2016vas,Davison:2016ngz,Yoon:2017nig,Narayan:2017hvh} and its holographic dual 2D gravity~\cite{Maldacena:2016upp}. Furthermore, the Schwarzian action features the one-loop exactness of the partition function~\cite{Stanford:2017thb} and its genus expansion demonstrates the random matrix behavior of 2d gravity~\cite{Saad:2019lba}.

It is remarkable that  JT gravity~\eqref{def: jt action} in the frame-like formulation can be described as $sl(2,\mathbb{R})$  BF theory~\cite{Fukuyama:1985gg,Chamseddine:1989wn,Chamseddine:1989yz} which is the reminiscence of the Chern-Simons formulation for the AdS$_3$ gravity~\cite{Achucarro:1986uwr,Witten:1988hc}. The BF formulation for  JT gravity has a great advantage in extending the space-time symmetry\footnote{Also See Ref.~\cite{Gomis:2021irw} as well as Ref.~\cite{Joung:2018frr} for their non-relativistic limits.  } simply by taking other gauge algebras instead of $sl(2,\mathbb{R})$: $sl(N,\mathbb{R})$ for the higher-spin extension~\cite{Alkalaev:2013fsa,Grumiller:2013swa,Alkalaev:2014qpa,Gonzalez:2018enk,Alkalaev:2019xuv,Alkalaev:2020kut}, $OSp(2,\mathcal N)$ for the supersymmetric extension~\cite{Astorino:2002bj,Cardenas:2018krd,Fan:2021wsb}. Another interesting extension is a color generalization of the spacetime which can be thought of as a multi-graviton theory. In~\cite{Alkalaev:2022szj} we introduced a colored version of JT  gravity with $su(N,N)$, which is analogous to the colored extension of 3D Einstein gravity via $su(N,N)$ Chern-Simons gravity~\cite{Gwak:2015vfb,Gwak:2015jdo,Joung:2017hsi}. Despite  no-go results and interpretation problems related to higher-dimensional multi-graviton constructions, in lower dimensions, BF and Chern-Simons theories of gravity can be safely color extended just  because of their pure topological nature. Nonetheless, there could be edge states interpreted as boundary ``multi-gravitons''.

In this paper, we begin with the $su(N,N)$ BF action for the colored JT gravity and impose a suitable asymptotic boundary condition leading to the boundary effective action. Since $SU(1,1) \approx SL(2,\mathbb{R})$ this construction can be considered as the color $N>1$ generalization of the Schwarzian action in the nearly-AdS$_2$~\cite{Maldacena:2016upp}. In the $SL(2,\mathbb{R})$ case, one can obtain the Schwarzian action at finite temperature by using the reparametrization transformation $\tau \to \tan \big({\pi \tau\over \beta}\big)$ from the zero-temperature one. However, for the case of the extended space-time symmetry like higher-spin or colored one, such a transformation to the finite temperature is not known. Hence, we consider an alternative way to derive boundary actions at finite temperature. At $N=1$  one may solve the asymptotic AdS$_2$ condition by the Gauss decomposition of the respective  $SL(2,\mathbb{R})$ group elements, which gives the zero temperature Schwarzian action~\cite{Narayan:2019ove,Saad:2019lba}. It was shown in~\cite{Valach:2019jzv} that the Iwasawa decomposition of $SL(2,\mathbb{R})$ group elements, instead of the Gauss decomposition, leads to the Schwarzian action at finite temperature. We propose the boundary effective action of the colored JT gravity at finite temperature which is built by using the Iwasawa decomposition of $SU(N,N)$ group elements.

The colored JT gravity accommodates interesting background solutions which could be referred to as rainbow-AdS$_2$ in which the colored gravitons do not vanish at asymptotic infinity.\footnote{This rainbow-AdS$_3$ vacua were investigated in the colored AdS$_3$ gravity~\cite{Gwak:2015vfb,Gwak:2015jdo,Joung:2017hsi}.} In the rainbow-AdS$_2$ background, the broken color gauge symmetry makes a part of colored gravitons unstable that resembles  partially-massless fields emerging via the  Higgs mechanism in the colored AdS$_3$ Chern-Simons gravity~\cite{Gwak:2015vfb,Gwak:2015jdo,Joung:2017hsi}. Furthermore, those modes also transform according to the modified asymptotic symmetry of the rainbow-AdS$_2$ background which we refer to as the asymptotic rainbow-AdS$_2$ symmetry.

The paper is organized as follows. In Section~\bref{sec: asympototic ads} we review the Schwarzian boundary action of the $sl(2,\mathbb{R})$ BF action. We discuss the isometry, the finite temperature extension and the respective smooth geometry. In Section~\bref{sec: color generalization} we build  the boundary effective action of the $su(N,N)$ BF action for the color generalization of JT gravity. The zero and finite temperature cases  are analyzed by means of the Gauss and Iwasawa group decompositions. In Section \bref{sec: quadratic action} we develop the semi-classical perturbative analysis of the boundary action and find  the form of the quadratic action which explicitly manifests the issues of (in)stability of the colored gravity. Also, we analyze the Lyapunov exponents and the bound on chaos. In Section~\bref{sec:rainbow} we introduce the notion of the rainbow-AdS$_2$ geometry and analyze the spectrum of fluctuations around the colored background. In Section~\bref{sec: discussions} we summarize our results and discuss future directions.

%%%%%%%%%%%%%%%%%%%%%%%%%%%%%%%%%%%%%%%%%%%%%%%%%%%%%%%%%
%%%%%%%%%%%%%%%%%%%%%%%%%%%%%%%%%%%%%%%%%%%%%%%%%%%%%%%%%
\section{%Asymptotic AdS solutions and Color-extension of Schwarzian 
Boundary Effective Action of $sl(2, \mathbb{R})$ JT gravity}
\label{sec: asympototic ads}
%%%%%%%%%%%%%%%%%%%%%%%%%%%%%%%%%%%%%%%%%%%%%%%%%%%%%%%%%
%%%%%%%%%%%%%%%%%%%%%%%%%%%%%%%%%%%%%%%%%%%%%%%%%%%%%%%%%

In this section we review  the uncolored JT gravity with emphasis on the points which are important in the subsequent construction of the color extension.

\subsection{Boundary reduction
of $sl(2,\mathbb R)$ JT gravity} 
\label{sec:schwarzian}

Let us discuss  the boundary  reduction of  $sl(2,\mathbb{R})$ BF formulation of JT gravity. 
The bulk part of the action  is given by
\begin{align}
\label{BF_N2}
    S_{BF}\,=\, \kappa \int_{\cM_2} \tr \big( \Phi\, F\big) \,,
\end{align}
where $F=dA + A\wedge A$ is a curvature 2-form,  a connection 1-form  $A$ and  
a dilaton  0-form $\Phi$  take values in $sl(2,\mathbb{R})$;
the trace $\tr$ stands for the invariant quadratic form of the gauge algebra.\footnote{The $sl(2,\mathbb{R})$ algebra basis elements are realized as
\be 
\label{LLL}
   	L_{-1}\,=\,\begin{pmatrix} 
	0 & -1 \\ 0 & 0
	\end{pmatrix}\,, 
	\quad 
	L_{0}\,=\,\begin{pmatrix} 
	{1\over 2} & 0 \\ 0 & -{1\over 2}
	\end{pmatrix}\,,  \quad  L_1\,=\,\begin{pmatrix} 
	0 & 0 \\ 1 & 0
	\end{pmatrix}\,, 
\ee
with the commutation relations $ [L_1, L_{-1}]\,=\,2 L_0$,\; $[L_0, L_{\pm1}]\,=\,\mp \,L_{\pm1}$. Then, $\tr$ is a matrix trace.
} 
The above BF action can be interpreted as the frame-like formulation of JT gravity upon identifying the components of $A$ with the zweibein and Lorentz spin connection and one component of $\Phi$ as the dilaton $\phi$ \cite{Fukuyama:1985gg,Chamseddine:1989wn,Chamseddine:1989yz}. The other two components of $\Phi$ play  the role of Lagrangian multiplier to impose  the torsion-free condition while the spin connection determines them algebraically through other fields. In this way, the $sl(2,\mathbb{R})$ BF action is reduced to JT  action~\eqref{def: jt action}, up to a boundary term, where the dimensionless coupling constant $\kappa$ is related to the Newton constant $G_N$ by $\kappa = (8\pi G_N)^{-1}$.

An appropriate boundary term is needed to be added to \eqref{BF_N2} since the two-dimensional manifold $\cM_2$ has an asymptotic boundary. $\cM_2$ is assumed to  have the topology of a half plane or an open disc  depending on whether the temperature of the theory is zero or not.  Both cases can be parameterized by the  spatial coordinate $r$ and  the Euclidean time coordinate $u$. When the temperature of the theory is finite, $u\in S^1$ ($u\simeq u+2\pi$), otherwise $u\in\mathbb R$. The asymptotic boundary $\partial \cM_2$ is identified with the $r=\infty$ curve\footnote{The other bound of $r$ is determined by the background solution. See the following discussions.}   parameterized by the coordinate $u$. Then, the coordinate parameterization of the fields reads $\Phi = \Phi(u,r)$ and $A = \{A_u(u,r), A_r(u,r)\}$.

The boundary term should be chosen together with an associated boundary condition so that the variational principle is well-defined in the presence of the boundary.\footnote{Let us remark that the choice of boundary terms and conditions is not unique. The full generality of all consistent choices  as well as the precise matching of such choices between the metric-like and frame-like formulations is yet to be  scrutinized.} In this paper, we  focus on the boundary term which is often used  in the frame-like formulation, 
\be 
    S_{bdy}\,=\,-{\kappa\over 2}\int_{\partial\mathcal M_2} 
    \tr(\Phi\,A)\,, \label{eq: boundary term sl2}
\ee 
along with its compatible boundary condition, 
\be 
    (\Phi+ \gamma\, A_u)\big|_{\partial\mathcal M_2}\,=\,0\,,  \label{eq: boundary condition sl2}
\ee
which altogether lead to the well-defined variational principle defined by the total action 
\be
\label{tot}
S_{tot} = S_{BF}+S_{bdy}\,,
\ee 
(see e.g. a discussion in   \cite{Mertens:2018fds,Saad:2019lba}). Here,  $\gamma$ is a real constant related to the boundary value of $\Phi$ which is determined by the boundary value of the dilaton field $\phi$ in the nearly-AdS$_2$ space \eqref{def: jt action}. From  \eqref{eq: boundary term sl2} and \eqref{eq: boundary condition sl2}  we can rewrite the boundary term as 
\be 
    S_{bdy}\,=\,    
    {\kappa \,\gamma\over 2 }
    \int_{\partial\mathcal M_2} {\rm d}u\;
 \tr(A_u^2)\,.\label{eq: boundary term sl2 2}
\ee 

In addition, we  demand the asymptotic AdS$_2$ condition in the expansion of the frame-like formulation~\cite{Brown:1986nw,Campoleoni:2010zq,Grumiller:2017qao} 
\begin{equation}
	\left.(A-A_{\text{\tiny AdS}})\right|_{\partial \mathcal{M}_2 }\,=\,\cO(1)\,, 
	\label{eq: asymptotic ads condition}
\end{equation}
where the  AdS$_2$ connection  $A_{\text{\tiny AdS}}$ is 
given by
\begin{align}
A_{\text{\tiny AdS}}= \big(e^r\,L_1+\cL_0\,e^{-r}\,L_{-1}\big)\,\rd u + L_0 \,\rd r\;.
\label{A_ads}
\end{align} 
Here, $\cL_0$ is a real constant which determines global properties  of the space (see section \bref{sec:back_sol}).

The Schwarzian action is obtained by evaluating the total action $S_{tot}$ on a particular  flat connection  supplemented with the  asymptotic AdS$_2$ condition \eqref{eq: asymptotic ads condition}. Such a connection has the gauge fixed form \cite{Banados:1994tn,Banados:1998gg}, 
\begin{equation}	
	A(u,r)\,=\,b^{-1}(r) \,a(u)\,{\rm d}u\,b(r) +b^{-1}(r)\,{\rm d}b(r)\,, 
	\qquad b(r)\,=\,e^{r\,L_0}\,,\label{eq: gauge choice sl2}
\end{equation}
where $a(u)$ is an $sl(2, \mathbb{R})$-valued function independent of the radial variable $r$ due to the flatness constraint satisfied by $A(u,r)$  in the bulk.  The  function $a(u)$ is further restricted by the asymptotic AdS$_2$ condition~\eqref{eq: asymptotic ads condition} and,  after a further gauge fixing, takes the form,
\begin{equation}
	a(u)\,=\,L_1+\mathcal L(u)\, L_{-1}\,=\,\begin{pmatrix}
	0 & -\cL(u)\\
	1 & 0\\
	\end{pmatrix}.   
	\label{AAdS}
\end{equation}
Here, as opposed to the AdS$_2$ connection
\eqref{A_ads},
the function $\mathcal L(u)$ does
depend on $u$. It is analogous to the boundary energy-momentum tensor arising in the asymptotic  analysis of the 3D Chern-Simons gravity. 
Since the BF action in the bulk vanishes after solving the flatness condition, 
the total action $S_{tot}$
%BF action with \ed{a} boundary term for nearly-AdS$_2$ \kostya{nearly-ads was not defined earlier in this section}
is reduced to the boundary action $S_{bdy}$ \eqref{eq: boundary term sl2 2} 
and we find
%Thus, from the Eq.~\eqref{AAdS} the total action becomes
%
\be 
    S_{tot}\,=\,-
    \kappa \,\gamma
    \int_{\partial\mathcal M_2}{\rm d}u\, \cL(u)\,.
    \label{Stot red}
\ee
Therefore, the task boils down
to determining the function $\cL(u)$.

\subsection{Background solutions}
\label{sec:back_sol}

Remark that Schwarzian action can be viewed as a partial on-shell action because
the other equation of motion, the covariant constancy condition of the dilaton, $d\Phi+[A,\Phi]=0$, is not yet imposed.  Imposing the latter  is equivalent to imposing the equation of motion of the Schwarzian theory  \cite{Maldacena:2016upp}. The covariant constancy condition reduces to
\be 
    \partial_u\phi(u,r)+[a(u),\phi(u,r)]=0\,,
    \qquad 
    \partial_r\phi(u,r)=0\,,
    \label{Cov Cons}
\ee 
where $\phi(u,r)=b(r)\,\Phi(u,r)\,b^{-1}(r)$. The second condition removes the $r$-dependence of $\phi$ that is consistent with  the boundary condition \eqref{eq: boundary condition sl2}\,: $\phi(u)+\gamma\,a(u)=0$\,. In the end, the first equation of 
\eqref{Cov Cons} yields the condition $\partial_u\,a(u)=0$ which has only constant solutions. It follows that  $\cL(u)=\cL_0$, and hence we recover the AdS$_2$ connection \eqref{A_ads} with a constant $\cL_0$.

We can rewrite the AdS$_2$ connections parameterized by different values of $\cL_0$ into the  metric,
\be 
    {\rm d} s^2=
    {\rm d}r^2+(e^r-\cL_0\,e^{-r})^2{\rm d}u^2\,,
\ee 
that leads to  solutions of  different topologies depending on the value of $\cL_0$. 
\begin{itemize}
\item[{\bf (1)}]
When $\cL_0>0$, we can parameterize $\cL_0=e^{2\,r_0}$ by some $r_0$ and find that
\be 
    {\rm d} s^2=
    {\rm d}r^2+\sinh^2(r-r_0)\,(2\,e^{r_0}\,{\rm d}u)^2\,.
\ee
The  metric is defined in the domain   $r\in (r_0,+\infty)$  and $r=r_0$ is the origin
which has a conical singularity unless  $u\simeq u+2\pi/(2\,e^{r_0})$.
The space has the topology of an open disk (a cigar-type geometry). The Euclidean AdS$_2$ solution corresponds
to the case $\cL_0=\frac14$ 
where the periodicity of $u$ is $2\pi$ and the origin is regular:
the global hyperboloid metric in the polar coordinate system. 

\item[{\bf (2)}] 
When $\cL_0=0$, we find the  Poincar\'e half-plane metric,
\be 
    {\rm d} s^2=
    {\rm d}r^2+e^{2r}\,{\rm d}u^2\,,
\ee
which is valid for $r, u\in (-\infty, +\infty)$. The space has  the topology of a half-plane. 

\item[{\bf (3)}] 
When $\cL_0<0$, we can parameterize it as $\cL_0=-e^{2\,r_0}$ and find
\be 
    {\rm d} s^2=
    {\rm d}r^2+\cosh^2(r-r_0)\,(2\,e^{r_0}\,{\rm d}u)^2\,,
\ee
which is valid for $r\in (-\infty, +\infty)$.
In this case, the time coordinate $u$ is not restricted:
it may belong to $\mathbb R$ or $S^1$
with any periodicity.
Note that the
hyperplane $r=-\infty$ is the other 
asymptotic boundary
disconnected from
the one at $r=+\infty$. 
The space
has the topology of 
a plane or a cylinder,
depending on
whether $u\in \mathbb R$ or $S^1$.
\end{itemize}

In this work, we will concern  the color generalization of the first two cases {\bf (1)} and {\bf (2)}, i.e. the two-dimensional hyperboloid with $\cL_0=\frac14$ and the Poincar\'e half-plane with $\cL_0=0$. With suitable regularity conditions at $r\to +\infty$ and $u\to \pm\infty$, the latter case also provides a good coordinate system on the two-dimensional hyperboloid, namely, the Poincar\'e coordinates. Therefore, if we formulate everything in a coordinate independent manner and the metric and dilaton fields are sufficiently regular,  the two cases cannot give  different results. In fact, the difference of the two cases arises from the asymptotic  boundary condition for $\Phi$ \eqref{eq: boundary condition sl2}, which is coordinate dependent. Due to this coordinate dependence, regular field configurations in one case  map to irregular ones in the other case, and vice versa, upon coordinate transformations.

\paragraph{Gauge function and holonomy.} When the background topology is that of a half-plane, see {\bf (2)}, the flatness condition assures that there exists a gauge function $\cG(r,u)$ 
such that  $A(r,u)=\cG^{-1}(r,u)\,{\rm d} \cG(r,u)$.
With the gauge choice~\eqref{eq: gauge choice sl2}, the connection $a(u)$ can be written as 
 $a(u)=g^{-1}(u)\,\dot g(u)$, where $\cG(r,u)=g(u)\, b(r)$ and  $\dot g(u) \,=\, \frac{\rm d}{{\rm d}u}g(u)$. We may also write the  action \eqref{Stot red}
 with $\Gamma\equiv\gamma\,L_{1}\in sl(2,\mathbb R)$ as
 \be 
 \label{coadj}
     S_{tot}\,=\,
    \kappa \,
    \int_{\partial\mathcal M_2}
    \tr(\Gamma\,g^{-1}\,{\rm d}g)\,,
 \ee 
which has the form of a coadjoint orbit action. However, $g(u)$ here  is not arbitrary but subject to the asymptotic AdS$_2$ condition \eqref{AAdS}.

When the background topology
is that of an open disk with a possible
conical singularity at the origin, see {\bf (1)}, 
the gauge function cannot be uniquely 
defined within a single coordinate patch.
Focusing on the background solutions,
we find the gauge function as
\be 
    g_{\text{\tiny AdS}}(u)=g_0\,\exp(a_{\text{\tiny AdS}}\,u)\,,
\ee 
where $g_0$ is a constant group element, and  the multi-valuedness of $g_{\text{\tiny AdS}}(u)$ is given by
the holonomy, 
\begin{align}
    \text{Hol}(A)\,=\, \cP \exp \bigg[ \oint A\bigg] \,\sim\, \exp \big(2\pi \,a_{\text{\tiny AdS}}\big) \,. 
\end{align}
If we restrict to the trivial holonomy, then the gauge function is single-valued
and defines a map from $S^1$ to $PSL(2,\mathbb R)$, provided  the flat connections were not
subject to the asymptotic AdS$_2$ condition.
On the other hand, the asymptotic AdS$_2$ condition restricts 
$PSL(2,\mathbb R)$ to its  one-dimensional subgroup $SO(2)$
generated by $a_{\textrm{\tiny AdS}}$ with
$\cL_0>0$\,. 
The solutions of the trivial holonomy  are parameterized as
\begin{align}
    \cL_0\,=\, { \nu^2\over 4}\;,\hspace{8mm}\text{where}\;\; \nu\in \mathbb{Z}_+\,.  
\end{align}
Note that we have used the projectivity of $PSL(2,\mathbb R)$.
The $\nu=1$ case corresponds to the AdS$_2$ background with $\cL_0={1\over4}$. Other connections with $\nu>1$ correspond to the geometry with conical surplus which we will not discuss in this paper.\footnote{This is related to the different coadjoint orbits considered in \cite{Mertens:2019tcm}.}

\subsection{Derivation of Schwarzian action}

Both the zero- and finite-temperature  Schwarzian actions can be obtained from  \eqref{Stot red}
by imposing the asymptotic AdS$_2$ condition \eqref{AAdS}.  In what follows  we review the relevant details.

\subsubsection{Zero-temperature Schwarzian}
\label{sec:zero_temp} 
 
 Using the Gauss decomposition of the gauge group element $g(u)$\,,
\be
	g(u)\,=\,\begin{pmatrix} 
	1 & 0 \\ f(u) & 1
	\end{pmatrix}
	\begin{pmatrix} 
	b^{-1}(u) & 0 \\ 0 & b(u)
	\end{pmatrix}
	\begin{pmatrix} 
	1 & -e(u) \\  0 & 1
	\end{pmatrix},\label{eq: gauss decomposition of g}
\ee
the boundary connection   $a = a(u)$ can be represented as follows 
\be
	a\,=\,g^{-1}\,\dot g
	\,=\,\begin{pmatrix}
	 -\frac{\dot{b}}{b}
	 +\frac{e\,\dot{f}}{b^2}
	 & \quad -\dot{e}  + 2\,\frac{\dot{b}}b\,e
	  -\frac{e^2\,\dot{f}}{b^2}\\
	\frac{\dot f}{b^2} &\frac{\dot{b}}{b}
	 -\frac{e\,\dot{f}}{b^2}
	\end{pmatrix}.
	\label{a eq}
\ee 
%where $\dot g \,=\, \frac{\rm d}{{\rm d}u}g(u)$.
%
By equating $a(u)$ given by \eqref{a eq}  with the one in \eqref{AAdS}, we obtain the  relations
\begin{equation}
\label{rels_sing}
\dot{f}\,=\,b^2\,, \qquad
 b^{-1}\,\dot{b}\,=\,e\,, \qquad
\dot{e}-e^2\,=\,\cL\,. 
\end{equation}
These allow us to express $\cL(u)$ in terms of $f(u)$, which turns out to coincide with the Schwarzian derivative% of $f(u)$: 
\be
	\cL\,=\,b^{-1}\,{\ddot b}-2\,(b^{-1}\,\dot b)^2
	\,=\,\frac12\bigg[\frac{\dddot f}{\dot f}-\frac32\left(\frac{\ddot f}{\dot f}\right)^2\bigg]
	\,\equiv \,\frac12\,{\rm Sch}\left(f,u\right).
	\label{Sch}
\ee
Finally, the total action on the nearly-AdS$_2$ background   is reduced to the Schwarzian action~\cite{Maldacena:2016upp}
\begin{align}
\label{Stotf}
    S_{tot}\,=\, - {\kappa \gamma\over 2}\int_{\partial\mathcal M_2} \rd u\; {\rm Sch}\left(f,u\right)\,.
\end{align}
This form is seemingly that of the Schwarzian action at zero temperature. However, we have not imposed any condition on the function $f(u)$ so far. Only when we restrict $f(u)$ to be a non-periodic function from $\mathbb{R}$  to $\mathbb{R}$, this will describe the Schwarzian theory at zero temperature.

%%%%%%%%%%%%%%%%%%%%%%%%%%%%%%%%%%%%%%%%%%%%%%%%%%%%%%%%%%%%%%%%%%%%%%
\subsubsection{Finite-temperature Schwarzian}
\label{sec: decomposition}
%%%%%%%%%%%%%%%%%%%%%%%%%%%%%%%%%%%%%%%%%%%%%%%%%%%%%%%%%%%%%%%%%%%%%%

From the Schwarzian derivative~\eqref{Sch} by using the compactifying coordinate transformation 
\be 
    f=\tan{\theta\over 2}\,,
    \qquad
    \theta\in (-\pi,\pi)\,,
    \label{f to th}
\ee 
one can get the finite temperature Schwarzian,
\be 
    \cL(u)\,=\, \frac12\left({\rm Sch}(\theta,u)+\frac12\,\dot\theta^2\right).
\ee 
However, this method 
of coordinate transformation 
cannot be immediately applied to Schwar\-zi\-an-like systems with extended symmetries,
as it would require a transformation beyond the diffeomorphism.\footnote{For example, the form of the higher-spin Schwarzian derivative at finite temperature is still an open question. Its perturbative expression was obtained in~\cite{Narayan:2019ove}. } Thus, in order to obtain the color-decorated Schwarzian at finite temperature from colored BF theory, 
we need to understand first
the reasoning behind the 
transformation \eqref{f to th}.
It is instructive to consider once again
background solutions in each cases.
\begin{itemize}
   
 \item 
In the zero-temperature case, $f$ is a function from $\mathbb R$ to $\mathbb R$. The function $f(u)=u$ is a background solution  and  the corresponding group element is given by  
\be 
    g(u)=\begin{pmatrix} 1&0\\ u& 1\end{pmatrix}.
\ee 
The two endpoints of $g(u)$ do not meet: $g(-\infty)\neq  g(\infty)$ and $g(-\infty)\neq  - g(\infty)$.
The curve goes  along an  $\mathbb R$ subgroup of 
$PSL(2,\mathbb R)$.

\item 
In the finite-temperature case, 
 $\theta(u)$
is a function from $S^1$ to $S^1$\,:
$\theta(u+2\pi)=\theta(u)+2\pi$.
The functions $\theta(u)=u$ or $f(u)=\tan\frac{u}2$ are
 a background solution of this kind and the corresponding group element is given by 
\be 
    g(u)=\begin{pmatrix} \cos\tfrac{u}2 & -\sin\tfrac{u}2\\ \sin\tfrac{u}2 & \;\;\; \cos\tfrac{u}2 \end{pmatrix}.
\ee 
The two endpoints of $g(u)$ coincide as elements of $PSL(2,\mathbb R)$. That is, up to 
the quotient by $\mathbb Z_2$ of $SL(2,\mathbb R)$ : $g(u+2\pi)= - g(u)$.  The curve goes  along a $SO(2)$ subgroup of  $PSL(2,\mathbb R)$ and it is non-contractible.

\end{itemize}

From the above, we conclude  that a finite temperature theory can be obtained from  the gauge functions which define non-contractible cycles of the group manifold, $PSL(2,\mathbb R)$ in the present  uncolored case.
For the non-contractibility,
the cycles should non-trivially wind
the maximal compact subgroup  
because the fundamental group 
of a non-compact group  is that of its maximal compact subgroup.
For this reason,
it will be convenient to use a group decomposition involving a maximal compact subgroup
and use its angle coordinate 
to parameterize the cycle.
The Iwasawa decomposition  precisely does this job. For any $g\in SL(2,\mathbb{R})$ it reads
\be
	g(\theta, c, e)\,=\,
	\begin{pmatrix}
	\cos {\theta\over 2} & -\sin {\theta\over 2} \\
	\sin {\theta\over 2} & \cos{\theta\over 2}	\end{pmatrix}
	\begin{pmatrix} 
	c^{-1} &  0 \\ 0 & c
	\end{pmatrix}
	\begin{pmatrix} 
	1 & -e\\ 0 & 1
	\end{pmatrix},  \label{eq: iwasawa decomposition}
\ee
where $c,e\in \mathbb R$ and 
$g(\theta+2\pi,c,e)\,=\,-g(\theta,c,e)\,\simeq\, g(\theta,c,e)$. 
Here,
we used 
the $\mathbb{Z}_2$ equivalence relation $\simeq$
in order to 
address an element of $PSL(2,\mathbb{R})=SL(2,\mathbb R)/\mathbb Z_2$.
In the same way as in Section~\bref{sec:zero_temp} with the Gauss decomposition, we evaluate the connection $a$  from the Iwasawa decomposition of $g(u)$~\eqref{eq: iwasawa decomposition} as 
\be
	a(u)\,=\, g^{-1}\,\dot g
	\,=\,\begin{pmatrix}
	-\frac{\dot c}{c}+\frac{e\,\dot\theta}{2 c^2} &\quad  -\dot e+2\frac{e\,\dot c}{c}-{1\over 2} c^2\,\dot\theta-\frac{e^2\,\dot \theta}{2c^2} \\
	\frac{\dot\theta}{2 c^2} & 	\frac{\dot c}{c}-\frac{e\,\dot\theta}{2 c^2}
	\end{pmatrix}.  \label{eq: gauge field sol sl2 finite temp}
\ee
Then, imposing the asymptotic AdS$_2$ condition~\eqref{AAdS} we find
the relations
\begin{equation}
\dot\theta\,=\, 2\,c^2\,,\qquad  c^{-1}\dot c\,=\,e\,, \qquad 
\dot e-e^2+\frac14\,\dot\theta^2\,=\,\cL\,, \label{eq: sol sl2 finite temp}
\end{equation}
and the action,
\be
	\label{f Sch}
\begin{alignedat}{2}
	S_{tot} \,=\,& -\kappa\gamma\int \rd u\;\cL(u)\,=\, -\kappa\gamma\int \rd u\big(  c^{-1}\,{\ddot c}-2\,(c^{-1}\,\dot c)^2+c^4\big) \cr
	\,=\,& - {\kappa\gamma\over 2} \int \rd u\; \bigg({\rm Sch}(\theta,u)
	+\frac12\,{\dot\theta}^2\bigg)\,. 
\end{alignedat}
\ee
Indeed, the Iwasawa decomposition successfully reproduces the finite temperature Schwarzian action.
This method of employing the Iwasawa decomposition can be applied straightforwardly to a more general case.

%%%%%%%%%%%%%%%%%%%%%%%%%%%%%%%%%%%%%%%%%%%%%%%%%%%%%%%%%%%%%%%%%
\subsection{Redundancy and Isometry}
\label{sec: isometry sl2}
%%%%%%%%%%%%%%%%%%%%%%%%%%%%%%%%%%%%%%%%%%%%%%%%%%%%%%%%%%%%%%%%%

In this section we discuss the issues of the redundancy in the Schwarzian action and the isometry of JT gravity within the BF formulation as a preparatory step for the analysis of the colored counterpart.

The redundancy arises from the fact that the map  
\be
\label{map_s}
s\;:\;\,g \mapsto a=g^{-1} \dot{g}
\ee 
is not injective: gauge functions $\tilde{g}\equiv k_0\, g$ with different constants $k_0\in SL(2,\mathbb{R})$ are all mapped to the same field $a(u)$. 
This implies that the resulting boundary action will have the corresponding symmetry which should be considered as an equivalence. In this sense, this symmetry is ``gauge'' rather than ``global''. In other words, if one considers the path-integral quantization of  the current theory one should not take into account the modes corresponding to this symmetry as it arises in the redefinition of the integration variable $a$ to $g$. In its turn, the redundancy in the gauge function $g$ induces a redundancy of the field variable $f(u)$ or $\theta(u)$ of the Schwarzian theory,
depending on the temperature of the theory.
In the zero-temperature case, 
using the Gauss decomposition  \eqref{eq: gauss decomposition of g} we can decompose $\tilde{g}=k_0\,g$ into $\tilde{f}, \tilde{c}$ and $\tilde{e}$. For the constant $k_0\in SL(2, \mathbb{R})$ given by
\be
\label{k0}
    k_0\,=\,\begin{pmatrix}
    c_4 & c_3\\
    c_2 & c_1\\
    \end{pmatrix},
    \quad c_1 c_4 - c_2 c_3 = 1\,,
\ee
we find that
\begin{align}
    \tilde{f}(u)\,=\, {c_1 f(u)+ c_2\over c_3 f(u) + c_4}\, .
    \label{eq: sl2 transformation}
\end{align}
This is the well-known $SL(2,\mathbb{R})$ invariance of the Schwarzian derivative which could be regarded as a redundancy (or a finite-dimensional ``gauge'' symmetry) rather than a physical (global) symmetry  of the Schwar\-zi\-an theory \eqref{Stotf}. 
In the finite-temperature case, using the Iwasawa decomposition \eqref{eq: iwasawa decomposition} of $\tilde{g}=k_0\,g$  we find
\begin{align}
\label{T_trans_sl2}
    \tan \tfrac{\theta}2  \quad\longmapsto \quad \tan \tfrac{\tilde\theta}2\,=\,{c_1 \tan {\theta \over 2} + c_2\over c_3 \tan {\theta \over 2} + c_4}\;.
\end{align}

In fact, the redundant description of the gauge connection $A(u,r)$ expressed in terms of $a(u)$ \eqref{eq: gauge choice sl2}  is related to the isometry of the AdS$_2$ background. The dynamical variable $f(u)$ of the Schwarzian theory can be 
viewed as the Nambu-Goldstone boson
resulting from
the symmetry breaking of diffeomorphism
by the AdS$_2$ background,
as explained in the metric-like formulation of JT gravity for the nearly-AdS$_2$~\cite{Maldacena:2016upp}.
Therefore, Schwarzian modes
correspond to
\be 
    \frac{\textrm{(Diffeomorphism)}}
    {\textrm{ (Isometry of AdS)}
    \times\textrm{(Boundary preserving diffeomorphism)}}\;,
\ee 
where we made the further quotient by (Boundary preserving diffeomorphism)
which is the genuine gauge symmetry.

\begin{figure}[h]
\centering

\begin{tikzpicture}[line width=0.6pt,scale=0.60]

\draw (0,0) node {$a_\ads$};
\draw [-stealth](0,-1) -- (0,-2);
\draw (0,-3) node {$a_\ads$};
\draw [-stealth](5,-3) -- (1.3,-3);

\draw (-5,0) node {$g_\ads$};
\draw [-stealth](-5,-1) -- (-5,-2);
\draw (-5,-3) node {$\tilde g_\ads$};
\draw [-stealth](-3.5,-3) -- (-1.7,-3);
\draw [-stealth](-3.5,0) -- (-1.7,0);
\draw [stealth-](7,-3) -- (10,-3);

\draw (6,0) node {$a$};
\draw [-stealth](6,-1) -- (6,-2);
\draw (6,-3) node {$a$};
\draw [-stealth](5,0) -- (1.3,0);
\draw [stealth-](7,0) -- (10,0);

\draw (11,0) node {$g$};
\draw [-stealth](11,-1) -- (11,-2);
\draw (11,-3) node {$\tilde g$};

\draw (-5.8,-1.5) node {$k_0$};
\draw (11.8,-1.5) node {$k_0$};
\draw (3,0.5) node {$h$};
\draw (3,-3.5) node {$h$};
\draw (-0.8,-1.5) node {$k$};
\draw (7,-1.5) node {$m$};
\draw (8.5,0.5) node {$s$};
\draw (8.5,-3.5) node {$s$};
\draw (-2.5,0.5) node {$s$};
\draw (-2.5,-3.5) node {$s$};

\end{tikzpicture}

\caption{A diagram of transformations underlying the relation between the redundancy and the isometry.  The maps \eqref{eq: parametrization of a sl2} and \eqref{iso_ads} define $h$ and $k$. A commutativity in the middle rectangle yields
$m = h^{-1} k h$ defining a stability transformation, $a =  m^{-1}\,a\, k + k^{-1} \dot{k}$, cf. \eqref{iso_ads}.
 The gauge group elements on the left and right parts of the diagram are transformed by left multiplications by $k_0$. The map $s$ assigns connections to group elements according to \eqref{map_s}.} 
\label{fig2}
\end{figure}

More explicit relations
can be shown by decomposing  the gauge field $A$ into
\be 
    A=H^{-1}\,\tilde A\,H+H^{-1}\,\rd H\,,
\ee 
where $H(r,u)=b^{-1}(r)\,h(u)\,b(r)$ is a function
determined by the Goldstone boson $h(u)$.
Although the gauge symmetry of $A$ 
is restricted to that of the boundary preserving 
diffeomorphism,
that of $\tilde A$
is not restricted because
it can be compensated by the transformation of $h(u)$.
Using the unrestricted gauge symmetry,
we can fix the flat connection $\tilde A$
as $A_{\rm AdS}$ given in \eqref{A_ads}
and express $a$ of \eqref{eq: gauge choice sl2} as
\begin{align}
    a \,=\, h^{-1}\,a_\ads\,h + h^{-1} \dot{h}\,. \label{eq: parametrization of a sl2}
\end{align} 
The above manifests that
there is an ambiguity in $h$, namely, $h\simeq k\,h$,
corresponding to the isometry of the background connection $a_\ads$:
\begin{align}
\label{iso_ads}
    k^{-1}\,a_\ads\, k + k^{-1} \dot{k} \,=\,a_\ads\,. 
\end{align}
By expressing $a_\ads$ using \eqref{map_s} as
\be 
    a_\ads=g_\ads^{-1}\,\dot g_\ads\,,
    \qquad 
    g_\ads=e^{a_\ads\,u}\,,
\ee 
we can parameterize $k(u)$ by a constant $k_0\in SL(2,\mathbb{R})$ as 
\begin{align}
    k\,=\,g_\ads^{-1}\, k_0\, 
    g_\ads\,. 
\end{align}
This form of the solution can be easily understood if one rewrites the equation \eqref{iso_ads} as the covariance constancy condition $\dot{k}+[a_\ads,k] =0$ which, therefore, defines the Killing symmetries of the background geometry.\footnote{On the other hand, recall that the dilaton $\Phi$ satisfies the covariant constancy condition which can be viewed as the gauge transformation preserving the form of the connection $A$. Then, using the boundary condition \eqref{eq: boundary condition sl2} and the form of the gauge fixed component $A_u$ of the background connection \eqref{eq: gauge choice sl2} one readily obtains the above Killing equation.} Also, for the gauge function $g_\ads$ the isometry is seen as the redundancy $g_\ads \simeq k_0 g_\ads  = \tilde g_\ads$.  When we express $a=g^{-1}\,\dot g$ \eqref{map_s}, the function $g$ is related to  the Goldstone boson $h$ by $g=g_\ads\,h$, and the ambiguity $h\simeq k\,h$ due to the isometry is mapped to the left invariance $g\simeq k_0\,g = \tilde g$\,. This consideration can be summarized in terms of a commutative diagram of maps depicted on Fig. \bref{fig2}.

%%%%%%%%%%%%%%%%%%%%%%%%%%%%%%%%%%%%%%%%%%%%%%%%%%%%%%%%%%%%%%%%%%%%%%%%%%%%
%%%%%%%%%%%%%%%%%%%%%%%%%%%%%%%%%%%%%%%%%%%%%%%%%%%%%%%%%%%%%%%%%%%%%%%%%%%%
\section{Boundary Effective Action of  the Colored JT Gravity}
\label{sec: color generalization}
%%%%%%%%%%%%%%%%%%%%%%%%%%%%%%%%%%%%%%%%%%%%%%%%%%%%%%%%%%%%%%%%%%%%%%%%%%%%
%%%%%%%%%%%%%%%%%%%%%%%%%%%%%%%%%%%%%%%%%%%%%%%%%%%%%%%%%%%%%%%%%%%%%%%%%%%%

Let us build  the color generalization of the $sl(2,\mathbb{R})$ BF theory for the nearly-AdS$_2$ which we have reviewed in the previous section. Noting the group isomorphism  $SL(2,\mathbb R)\simeq SU(1,1)$ we can  consider $su(N,N)$ BF theory for the colored JT gravity. In \cite{Alkalaev:2022szj} we developed the $su(N,N)$ BF formulation and its relation to the second-order formulation of the colored JT gravity. In what follows  we  study how to reduce the colored JT gravity to the color-decorated Schwarzian theory along the lines of the standard $SL(2,\mathbb{R})$ case. 

%Hence, let us introduce the basic materials of this representation before going to the main analysis.

To implement the reduction in  the $su(N,N)$ case it is convenient to work in  a representation isomorphic to the fundamental one which at $N=1$ reduces to the fundamental representation of $SL(2,\mathbb R)$. Using the fundamental representation the isomorphism $SL(2,\mathbb R)\simeq SU(1,1)$ can be achieved by the unitary similarity transformation,
\be
\label{isom}
	A\in SL(2,\mathbb R)\,\longleftrightarrow\,
	B\,=\, U^{-1}\,A\,U\in SU(1,1)\,,
	\qquad
	U\,\equiv\,\frac1{\sqrt{2}}\,\begin{pmatrix} 1 & -i \\ -i & 1 \end{pmatrix}.
\ee
The similar trick  can be used in  the $SU(N,N)$ case. To this end, we recall the standard matrix parameterization
\be
B\in SU(N,N): \qquad B^\dagger \begin{pmatrix} \bm I & 0 \\ 0 & -\bm I \end{pmatrix} B = \begin{pmatrix} \bm I & 0 \\ 0 & -\bm I \end{pmatrix},
\qquad
\det B\,=\,1\,, 
\ee 
where $\bmI$ is $N\times N$ unit matrix. Then, the similarity transformation   
\be
A\,=\,U\,B\,U^{-1}\,,
	\quad
	U\,\equiv\,\frac1{\sqrt{2}}\,\begin{pmatrix} \bm I & -i\,\bm I \\ -i\,\bm I & \bm I \end{pmatrix}, 
	\quad 
	U^\dagger \,=\, U^{-1}\;,
	\label{unitary}
\ee 
yields the following matrix representation:
\be
A\in SU(N,N): \quad	A^\dagger\,\begin{pmatrix} \bm 0 & -\bm I \\ \bm I & \bm 0 \end{pmatrix}
	A\,=\,\begin{pmatrix} \bm 0 & -\bm I \\ \bm I & \bm 0 \end{pmatrix},
	\qquad
	\det A\,=\,1\,,
	\label{SU(N,N)}
\ee
which is isomorphic to the fundamental one by construction. Such a representation  can be viewed as a color extension of $SL(2,\mathbb R)$ understood as replacing entries of $SL(2, \mathbb{R})$ $2\times 2$  matrix with  $N\times N$ blocks representing colors. An infinitesimal expansion  $A=I_{2N}+\epsilon X +\cO(\epsilon^2)$, where $I_{2N}$ is $2N\times 2N$ unit matrix, provides the corresponding representation
of the algebra $su(N,N)$ as
\be
	\tr X=0\,,\qquad
	X^\dagger=\begin{pmatrix} \bm 0 & -\bm I \\ \bm I & \bm 0 \end{pmatrix}\,X\,\begin{pmatrix} \bm 0 & -\bm I \\ \bm I & \bm 0 \end{pmatrix}.
	\label{su(N,N)}
\ee
Any $X$ satisfying the condition~\eqref{su(N,N)} can be expressed as 
\be
\label{X_par}
	X\,=\,\begin{pmatrix} -\bm\beta+i\,\bm \delta & -\bm \epsilon \\ \bm \varphi & \bm \beta+i\,\bm \delta \end{pmatrix}, 
\ee
where $\bm \beta, \bm \delta, \bm \epsilon, \bm \varphi$ are Hermitian matrices and  $\tr\bm \delta=0$.
Exponentiations of the generators corresponding to $\bm \beta, \bm \delta, \bm \epsilon, \bm \varphi$ can be given as, respectively, 
\be
\label{bdfe}
	\begin{pmatrix} \bm b^{-1} & \bm 0 \\ \bm 0& \bm b \end{pmatrix},
	\qquad
	\begin{pmatrix} \bm d & \bm 0 \\ \bm 0 & \bm d \end{pmatrix},
	\qquad
	\begin{pmatrix} \bm I & -\bm e \\ \bm 0 & \bm I \end{pmatrix},
	\qquad
	\begin{pmatrix} \bm I & \bm 0 \\ \bm f & \bm I \end{pmatrix},
\ee 
where matrices $\bm b, \bm e, \bm f$ are  Hermitian, whereas $\bm d \in SU(N)$.

Yet another useful parameterization of the algebra $su(N,N)$ is given by   the following decomposition  
\begin{equation}
\label{suNN_dec}
	su(N,N)\,\simeq\, 
	\big[su(1,1)  \otimes \bmI \big]\,\oplus\,
	\big[I\otimes su(N)\big]\,\oplus\, 
	\big[su(1,1) \otimes su(N)\big]\;,
\end{equation}
where $I$ is  $2\times 2$ identity matrix. An arbitrary element $M\in su(N,N)$ can be represented as
\be
\label{suNN}
M \,=\, \cL^m (L_m\otimes \bmI) +i\cJ^A  (I \otimes \bT_A)
+ \cK^{m,A}(L_m \otimes \bT_A)\;,
\ee
where $\cL^m$, $\cJ^A$, $\cK^{m,A}$, with indices $m=0,\pm1$ and $A = 1,..., N^2-1$, are real parameters, while matrices  $(I, L_m)$ are $u(1,1)$ basis elements \eqref{LLL}, and Hermitian $N\times N$ matrices $(\bmI, \bT_A)$  are  $u(N)$ basis elements (for more details see, e.g. \cite{Joung:2017hsi,Alkalaev:2022szj}).  The latter satisfy the product relation
\be
\label{prod_T}
\bT_A\bT_B\,=\, {1\over N} \delta_{AB} \,\bmI + \big({g_{AB}}^C + i {f_{AB}}^C\big)\,\bT_C\,, 
\ee
where $g_{ABC}$ and $f_{ABC}$ are totally (anti-)symmetric real constants. The basis elements of $su(N,N)$ read  
\be
\label{basis}
G_m \,=\, L_m\otimes \bmI\;,
\qquad
G_A \,=\, I \otimes  \bT_A \;,
\qquad
G_{m,A} \,=\, L_m \otimes \bT_A\;,
\ee
where $G_m$ and $G_A$ form the subalgebras $sl(2,\mathbb{R})$ and $su(N)$, respectively.

%%%%%%%%%%%%%%%%%%%%%%%%%%%%%%%%%%%%%%%%%%%%%%%%%%%%%%%%%%%%%%%%%%%%%%%%%%%%
\subsection{Colored Schwarzian Action at Zero Temperature}
\label{sec:zero_temp color}
%%%%%%%%%%%%%%%%%%%%%%%%%%%%%%%%%%%%%%%%%%%%%%%%%%%%%%%%%%%%%%%%%%%%%%%%%%%%

As a colored  extension of the $su(1,1)$ BF theory \eqref{BF_N2}, let us consider the $su(N,N)$ BF theory defined by the action 
\begin{align}
    S_{BF}\,=\, {\kappa\over N} \int_{\cM_2} \tr \big( \Phi\, F\big)\,, \label{eq: bulk bf action color}
\end{align}
where $F=dA + A\wedge A$ is the curvature 2-form,  the connection 1-form $A$ and  
the dilaton 0-form $\Phi$  now take values in $su(N,N)$, the trace $\tr$ stands for the invariant quadratic form of the gauge algebra. 

Using \eqref{suNN} and \eqref{basis} the BF $p$-forms $\Upsilon := (\Phi,A,F)$ can be represented as  
\be
\label{spectrum}
\Upsilon\,=\,\Upsilon^m G_m + i\Upsilon^A  G_A+\,\Upsilon^{m,A}G_{m,A}\,.
\ee
The component expansion \eqref{spectrum} defines the spectrum of the colored JT gravity \cite{Alkalaev:2022szj}:  

\begin{itemize}
\item  $\Upsilon^m$ are in the adjoint of  $sl(2,\mathbb{R})\approx su(1,1) \subset su(N,N)$ subalgebra. This color-singlet sector describes JT  dilaton gravity.  

\item  $\Upsilon^A$ are in the $su(N)\subset su(N,N)$ adjoint representation. This is the $su(N)$ BF sector.

\item  $\Upsilon^{m,A}$ are in the tensor product of $sl(2,\mathbb{R})$ and $su(N)$ adjoint representations. They form $su(N)$ adjoint multiplet of colored gravitons and dilatons.   

\end{itemize}

\noindent In total, there are  JT graviton and dilaton, $(N^2-1)$ colored gravitons and $(N^2-1)$ colored dilatons, $su(N)$ gauge fields and  matter fields forming the $su(N)$ adjoint multiplet. We refer to them as (colored) ``spin-2'' and ``spin-1'' modes, respectively.

\paragraph{Gauge and asymptotic conditions.} As in Section~\bref{sec: asympototic ads}, we can choose a gauge   
\begin{equation}	
	A(r,u)\,=\,b^{-1}(r) \,a(u)\,{\rm d}u\,b(r) +b^{-1}\,{\rm d}b(r)\,,
	\qquad b(r)\,=\,e^{r\,L_0\otimes \bmI}\,. \label{eq: color FG gauge}
\end{equation}
The difference with the $sl(2, \mathbb{R})$ case \eqref{eq: gauge choice sl2} is in the form of the gauge function $b(r)$ which is now exponentiation of  the $sl(2, \mathbb{R})$ generator $G_0=L_0 \otimes \bmI$ which is trivially color-extended  $sl(2,\mathbb{R})$ generator $L_0$ \eqref{basis}.  Noting that the theory admits a consistent truncation to the pure gravitational sector $sl(2,\mathbb{R})\subset su(N,N)$ the global AdS$_2$ solution is given by the connection 
\begin{align}
    a_\ads\,=\,(L_1 + \cL_0\, L_{-1})\,\otimes \bmI \,.
    \label{ads_sunn}
\end{align}
It can be interpreted as describing the background geometry with all other fields of the theory   vanishing. Using the representation \eqref{suNN} the asymptotic AdS$_2$ condition~\eqref{eq: asymptotic ads condition} can be translated into the following form of the connection $a(u)$ after we fix a suitable residual gauge symmetry, 
\be
\label{as_1}
\begin{alignedat}{2}
	a(u)\,=\,&  (L_1\otimes \bm{I})+\mathcal{L}(u)\,(L_{-1} \otimes \bm I)
              +i\,\mathcal{J}_A(u)\,(I \otimes \bm \bT^A)
		+ \mathcal{K}_A(u)\,(L_{-1}  \otimes \bm \bT^A)\\
	\,=\,&  L_1\otimes\bm{I}+i\,I\otimes \bm{\mathcal J}(u) 
	+L_{-1}\otimes \bm{\mathcal L}(u) \,,  
\end{alignedat}
\ee
where we  defined quantities 
\be
	\bm{\mathcal J}(u)\,=\,\mathcal J_A(u)\,\bm \bT^A\,, \qquad 
	\bm{\mathcal L}(u)\,=\,\mathcal L(u)\,\bm I + \mathcal K_A(u)\,\bm \bT^A\,, 
	\label{JL}
\ee
which are  traceless and traceful $N\times N$ Hermitian matrices, respectively. It follows that in the representation \eqref{X_par} the connection   $a(u)$ takes the  form
\be
	a(u)\,=\,\begin{pmatrix} 
	i\,\bm\cJ & -\bm\cL \\ \bm I & i\,\bm \cJ
	\end{pmatrix}. 
	\label{cAAdS}
\ee
Now, we want to identify a group element $g(u)\in SU(N,N)$ such that
 $a(u)\,=\,g^{-1}(u)\,\dot{g}(u)$ \eqref{map_s}. Using the representation \eqref{bdfe} the element $g(u)$ can be represented through the Gauss-like  decomposition as\footnote{\label{f:gauss}Note
that here we do not use  the Gauss
decomposition $L D U$, where $D$ is the diagonal matrix, and $L$ and $U$  are the nilpotent subgroups corresponding to the lower and upper triangular matrix, respectively.}
\be
	g(u)=\begin{pmatrix} 
	\bm I & \bm 0 \\ \bm f & \bm I
	\end{pmatrix}
	\begin{pmatrix} 
	\bm b^{-1} & \bm 0 \\ \bm 0 & \bm b
	\end{pmatrix}
	\begin{pmatrix} 
	\bm d& \bm 0 \\ \bm 0 & \bm d
	\end{pmatrix}
	\begin{pmatrix} 
	\bm I & -\bm e \\ \bm 0 & \bm I
	\end{pmatrix},
	\label{g decomp}
\ee
where matrices $\bm b, \bm e, \bm f$ are Hermitian 
and $\bm d \in SU(N)$. Then, we find that
\be
	a(u)\,=\,g^{-1}\,\dot g
	\,=\,\begin{pmatrix}
	\bm p & \bm q \\
	\bm r & \bm s
	\end{pmatrix}, \label{eq: sol a decomposition}
\ee
with 
\be
\label{eq: sol a decomposition_21}
\begin{alignedat}{2}
	\bm p\,=\,&
	 \bm d^{-1}\,\dot{\bm d} -\bm d^{-1}\,\dot{\bm b}\,\bm b^{-1}\,\bm d
	 +\bm e\,\bm d^{-1}\,\bm b^{-1}\,\dot{\bm f}\,\bm b^{-1}\,\bm d\,,  \\
	 \bm q\,=\,&-\dot{\bm e} -[ \bm d^{-1}\,\dot{\bm d},\bm e] + \bm d^{-1}\,\dot{\bm b}\,\bm b^{-1}\,\bm d\,\bm e
	 +\bm e\,\bm d^{-1}\,\bm b^{-1}\,\dot{\bm b}\,\bm d
	  -\bm e\,\bm d^{-1}\,\bm b^{-1}\,\dot{\bm f}\,\bm b^{-1}\,\bm d\,\bm e\,,  \\
	 \bm r\,=\,&
	  \bm d^{-1}\,\bm b^{-1}\,\dot{\bm f}\,\bm b^{-1}\,\bm d\,, \\
	  \bm s\,=\,&  \bm d^{-1}\,\dot{\bm d} +\bm d^{-1}\,\bm b^{-1}\,\dot{\bm b}\,\bm d
	                          - \bm d^{-1}\,\bm b^{-1}\,\dot{\bm f}\,\bm b^{-1}\,\bm d\,\bm e\,. 
\end{alignedat}
\ee
%
%
%Other decomposition
%
%\begin{align}
%	\bm p\,=\,&
%	 - \dot{\bm b}\,\bm b^{-1} + \bm b\,\bm d^{-1} \,\dot{\bm d}\,\,\bm b^{-1}
%	 +\bm e\,\bm b^{-1}\,\bm d^{-1}\,\dot{\bm f}\,\bm d\,\bm b^{-1}\,,  \nn
%	 \bm q\,=\,&-\dot{\bm e} - \bm b\,[  \bm d^{-1}\,\dot{\bm d} ,\,\bm b^{-1}\,\bm e\, \,\bm b^{-1}]\,\bm b + \dot{\bm b}\,\bm b^{-1}\,\bm e
%	 +\bm e\, \bm b^{-1}\,\dot{\bm b}
%	  -\bm e\,\bm b^{-1}\,\bm d^{-1}\,\dot{\bm f}\,\bm d\,\bm b^{-1}\,\bm e\,,  \nn
%	 \bm r\,=\,&
%	  \bm b^{-1}\,\bm d^{-1}\,\dot{\bm f}\,\bm d\,\bm b^{-1}\,, \nn
%	  \bm s\,=\,&  \bm b^{-1}\,\dot{\bm b} +\bm b^{-1}\,\bm d^{-1}\,\dot{\bm d}\,\bm b
%	  - \bm b^{-1}\,\bm d^{-1}\,\dot{\bm f}\,\bm d\,\bm b^{-1}\,\bm e\,. 
%\end{align}
%
%
By equating \eqref{eq: sol a decomposition} with \eqref{cAAdS} after some algebra  we obtain the  relations
\begin{align}
\label{rels_col}
 \dot{\bm f}\,=\,&\bm b^2\,,\\
\bm d^{-1}\,\bm b^{-1}\,\dot{\bm b}\,\bm d+\bm d^{-1}\,\dot{\bm d} \,=\,&i\,\bm\cJ+\bm e\,,\label{2eq}\\
 -\bm d^{-1}\,\dot{\bm b}\,\bm b^{-1}\,\bm d+\bm d^{-1}\,\dot{\bm d} \,=\,&i\,\bm\cJ-\bm e\,,\label{3eq}\\
 \dot{\bm e}-\bm e^2+i\,[\bm \cJ,\bm e]\,=\,&\bm \cL\,.\label{4eq}
\end{align}
Having in total four matrix differential equations for six independent matrix functions we find out that  $\bm b$, $\bm e$, $\bm\cJ$, $\bm \cL$ can be algebraically expressed in terms of $\bm d, \bm f$ and  their derivatives. What matters for us are the following traced quantities,
\be
	\tr(\bm \cL)\,=\,\frac{d}{du} \tr{\bm e}- \tr(\bm e^2)
	\,=\,\tr \left[\frac{d}{du}(\bm b^{-1}\dot{\bm b})
	-\left(\frac{\bm b^{-1}\dot{\bm b}+\dot{\bm b}\,\bm b^{-1}}2\right)^2\right], \label{eq: sol l color}
\ee
and
\be
	\tr(\bm \cJ^2)\,=\,-\,\tr\left[(\bm d^{-1}\,\dot{\bm d})^2
	+\dot{\bm d}\,\bm d^{-1}\left(\bm b^{-1}\dot{\bm b}-\dot{\bm b}\,\bm b^{-1}\right)
	+\left(\frac{\bm b^{-1}\dot{\bm b}-\dot{\bm b}\,\bm b^{-1}}2\right)^2\right], 
	\label{eq: sol j2 color}
\ee
which directly  follow from equations \eqref{2eq}-\eqref{4eq}.

\paragraph{Identifications via cyclic groups.} So far we have  considered $SU(N,N)$ as the color generalization of $SL(2,\mathbb R)\simeq SU(1,1)$, 
but the theory depends, in fact,  only on the Lie 
algebra structure $su(N,N)$
and hence $PSU(N,N)$ will be sufficient because there are no matter fields in addition to the connection which may transform non-trivially under the center group $Z[SU(N,N)]$.
If we consider $SU(N,N)$ instead of $PSU(N,N)$,
the action will acquire
an integer factor corresponding 
to the order of the center group,
which can be simply
absorbed into the coupling constant $\kappa$.
Remark that 
$Z[SU(N,N)]
=\{e^{2\pi\,i\frac{n}{2N}}\,I_{2N}\,|\, n=0,..., 2N-1 \}\simeq \mathbb Z_{2N}$
and, hence, $PSU(N,N)=SU(N,N)/\mathbb Z_{2N}$. A few comments are in order.  

\begin{itemize}

\item Since $\mathbb Z_2\subset\mathbb Z_{2N}$, we can first  factor  out  $\mathbb Z_2=\{\pm I_{2N}\}$  by disregarding the overall sign in $\bm b$ because $g \simeq -g$ in this case and $\bm b$ enters the decomposition  \eqref{g decomp} only through the block-diagonal second factor. This resolves the sign ambiguity in the solutions of $\bm b^2=\dot{\bm f}$ \eqref{rels_col}. However, there still remain multiple solutions for $\bm b$ even after factoring out  $\mathbb Z_2$ and, hence, there are multiple branches of solutions of this matrix equation.\footnote{Suppose that the matrix $\dot{\bm f}$ is diagonalizable with eigenvalues $\lambda_1,\ldots, \lambda_{N}$. Then $\bm b$ is a diagonal matrix with entries $\pm \sqrt{\lambda_i}$ up to an overall sign. Therefore, we find as many solutions as possible sign choices. The number depends on the degeneracy: for the maximally degenerated case there are  $\lceil(N+1)/2\rceil$ possibilities.} 

\item After the initial $\mathbb Z_2$ factorization, we still have $\mathbb Z_N = \{e^{2\pi\,i\frac{k}{N}}\,I_{2N}\,|\, k=0,..., N-1\}$ because $\mathbb Z_2$ naturally acts on $\mathbb Z_{2N}$ so that $\mathbb Z_{N} =\mathbb Z_{2N}/\mathbb Z_2$.  Obviously, these $\mathbb Z_N$ phase factors  are not associated with the different solution branches of the equation $\bm b^2=\dot{\bm f}$.

\item Instead,  the $\mathbb Z_N$ can be quotiented rather from the $SU(N)$ element $\bm d$ which enters the decomposition \eqref{g decomp} only through the block-diagonal third factor. The 
$\mathbb Z_N$ quotient will lead to $PSU(N)$ instead of $SU(N)$ for
the spin-1 part of the BF theory. 

\end{itemize}

\paragraph{Boundary effective action.} As in Section~\bref{sec: asympototic ads}, the bulk BF action~\eqref{eq: bulk bf action color} vanishes with the above asymptotic AdS$_2$ solutions \eqref{eq: color FG gauge} and \eqref{ads_sunn}. Hence, the effective action for the asymptotic AdS$_2$ solutions comes from the boundary term analogous to \eqref{eq: boundary term sl2},
\be 
    S_{bdy}\,=\,-{\kappa\over 2N}\int_{\partial\mathcal M_2} 
    \tr(\Phi\,A)\,,
\ee 
together with the boundary condition~\eqref{eq: boundary condition sl2}. Using \eqref{eq: color FG gauge} and \eqref{cAAdS} along with the relations \eqref{eq: sol l color} and \eqref{eq: sol j2 color} we obtain the boundary effective action for the colored JT gravity as
\be
    S_{tot}\,=\,\frac{\kappa\,\gamma}{2N} \int_{\partial \mathcal M_2}\rd u\,\tr(A_u^2)\,=\,-
    \frac{\kappa\,\gamma}{N}\int_{\partial \mathcal M_2}
    \rd  u\,\tr(\bm\cL+\bm\cJ^2)\,,\label{eq: boundary action zero temp}
\ee 
where $\tr(\bm\cL+\bm\cJ^2)$ is given by
\be 
	\tr(\bm\cL+\bm\cJ^2)\,=\,
	\tr \left[\bm b^{-1}\ddot{\bm b}-2(\bm b^{-1}\dot{\bm b})^2
	-(\bm d^{-1}\,\dot{\bm d})^2
	-\dot{\bm d}\,\bm d^{-1}\left(\bm b^{-1}\dot{\bm b}-\dot{\bm b}\,\bm b^{-1}\right)
\right]. 
\label{mat Sch}
\ee
Note that the boundary action~\eqref{eq: boundary action zero temp} is a functional of $\bm b$ and $\bm d$, while $\bm b^2$ itself is given by $\bm f$ in \eqref{rels_col}. A few comments are in order.

Comparing the above result with the usual Schwarzian action~\eqref{Sch}, we find that the variable  $b$ in  \eqref{Sch} is generalized to the 
$N\times N$ Hermitian matrix $\bm b$ where
the trace part can be viewed as the ``JT graviton'' whereas the traceless parts are the colored spin-$2$ modes. 

On the other hand, the group element $\bm d \in (P)SU(N)$ can be viewed as a  variable of the boundary reduced $(P)SU(N)$ BF theory. Indeed, turning off the gravitational part  by  setting  $\bm b = \bmI$ we end up with a particle  Lagrangian 
\be 
\tr(\bm\cJ^2)\,=\,-\tr \left(\bm d^{-1}\,\dot{\bm d}\right)^2\,,
\label{sigma_mod}
\ee
treated as that of $1d$  non-linear $\sigma$-model on the $(P)SU(N)$ group manifold. 

%\rmv{Now, we can turn off the spin-1 modes by setting $\bm d = \bmI$ in which  case  $\tr \bmcJ^2 = 0$. Besides the above natural identifications generated by the cyclic subgroups, we find a novel subtlety:} 
The above natural identifications generated by the cyclic subgroups, we find a novel subtlety: as opposed to the equation $b^2=\dot f$ \eqref{rels_sing}, the matrix equation $\bm b^2=\dot{\bm f}$ \eqref{rels_col} is not direct to solve. Nonetheless, it turns out that the boundary action \eqref{eq: boundary action zero temp} which is now given by the only contribution $\tr(\bm\cL)$ can be cast into  a simple form in terms of $\bm f$,
\be 
\label{col_schw}
    \tr(\bm\cL)
    \,=\,\frac12\,\tr\left[\dot {\bm f}^{-1}\,\dddot{\bm f}
    -\frac32\,(\dot{\bm f}^{-1}\ddot{\bm f})^2\right]\,.
\ee
This is the color generalization of the Schwarzian derivative for JT graviton and colored spin-$2$ modes. A few remarks are in order.

First, the matrix Schwarzian derivative in Eq.~\eqref{col_schw} was studied for $Sp(2N,\mathbb R)$ case in Refs.~\cite{AFST_1993_6_2_1_73_0,Ovsienko_2004}  as the higher-dimensional analogue of the Schwarzian derivative\footnote{We thank the referee for pointing this out.}. In this work, the $SU(N,N)$ Schwarzian~\eqref{col_schw} has been obtained from the colored JT gravity. Accordingly, the invariance of the matrix Schwarzian action under a $SU(N,N)$ transformation, which was proven in Refs.~\cite{AFST_1993_6_2_1_73_0,Ovsienko_2004} for the $Sp(2N,\mathbb R)$ case, naturally follows from the physical condition (the isometry of the $AdS_2$ background) of the colored JT gravity (e.g. see Section~\ref{sec: isometry of colored gravity}). Also note that the Schwarzian derivatives for higher-rank groups have been considered perviously in the literature, see e.g. \cite{Marshakov:1989ca,Li:2015osa} for $SL(N,
\mathbb R)$ groups in the context of $W$-gravity. 

Second, we may consider a restriction of the $su(N,N)$ gauge algebra of BF theory
to a certain subalgebra by reducing the  spin-1 gauge algebra,
according to the construction \cite{Vasiliev:1986qx,Konstein:1989ij}.
For instance, if we consider the $so(N)$ color instead of $su(N)$,
the $su(N,N)$ gauge algebra will be replaced by $sp(2N,\mathbb R)$ with
symmetric matrix $\bm f$ and $\bm d\in SO(N)$.
For even $N$, we can also consider the $usp(N/2)$ color
which requires the $so(N,N)$ gauge algebra with antisymmetric matrix $\bm f$
and  $\bm d\in U\!Sp(N/2)$.

Let us now consider small fluctuations described by the Schwarzian \eqref{col_schw}. Indeed, when the traceless  part of $\bm f$ is sufficient small, we can do a perturbative analysis.\footnote{Here,
we focus
the branch connected 
to the solution with $\bm b=\bm I$. Note that there 
exist multiple branches
satisfying $\dot{\bm f}=
\bm b^2$\,.}
We separate the trace and traceless parts as
\be 
    \bm f=f\,\bm I+\bm k\,,
    \qquad 
    \tr\bm k=0\,,
    \label{singlet decmp}
\ee 
and solve the relation $\bm b^2=\dot{\bm f}$
perturbatively in $\bm k$.
In the end, the action 
can be expanded in the powers of $\bm k$
or equivalently in
the powers of $\bm\alpha=\dot f^{-1}\,\dot{\bm k}$.
Using 
\begin{align} 
    \bm b^{-1}\,\dot{\bm b}
    -\dot{\bm b}\,\bm b^{-1}
    \,=\,&-\frac14 [\bm\alpha,\dot{\bm\alpha}]+\cO(\bm\alpha^3)\,,\\
    \bm b^{-1}\,\dot {\bm b}
    +\dot{\bm b}\,\bm b^{-1}
    \,=\, &\dot f^{-1}\,\ddot f
    +\dot{\bm \alpha}-\frac14\,\{\bm\alpha, \dot{\bm\alpha}\}+\cO(\bm\alpha^3)\ ,
\end{align}
we find
\be
\begin{alignedat}{2}
    -\tr(\bm \cL+\bm\cJ^2)
    \,=\,&-\frac N2\left( \dot f^{-1}\,\dddot f-\frac32 (\dot f^{-1}\,\ddot f)^2\right)
    +\frac12\tr\left(
    \bm\alpha\,\ddot{\bm\alpha}+
    \frac32\dot{\bm\alpha}^2
    \right)
    +\tr\left(\bm d^{-1}\,\dot{\bm d}\right)^2\\
    &-\,\frac14\,\dot f^{-1}\,\ddot f\,\tr(
    \bm\alpha\,\dot{\bm\alpha})
    -\frac14\,\tr\left(\dot{\bm d}\,\bm d^{-1}
    [\bm\alpha,\dot{\bm\alpha}]\right)
    +\cO(\bm\alpha^3)\,.
    \label{k exp}
\end{alignedat}
\ee
The first line 
consists of the Schwarzian action of the singlet graviton \eqref{Sch}-\eqref{Stotf}, a quadratic action of $\bm \alpha$
and the action for a particle on the group manifold $(P)SU(N)$ corresponding to the spin-1 mode \eqref{sigma_mod}.
The second term contains the action of 
free colored gravitons.
Indeed, one can check that the kinematic nature 
of the singlet graviton $f$ and the colored gravitons $\bm k$  are the same: they
have the same kinetic terms up to total derivatives, 
\be 
    -\bigg(\dot f^{-1}\,\dddot f-\frac32\,(\dot f^{-1}\,\ddot f)^2\bigg)
    \,=\,-\dddot{h} + \left(\dot h\,\dddot{h}
    +\frac32\,\ddot{h}^2\right)+\cO(h^3)\,,
\ee
and 
\be 
    \tr\left(
    \bm\alpha\,\ddot{\bm\alpha}+
    \frac32\dot{\bm\alpha}^2
    \right)
    \,=\,\tr\left(
    \dot{\bm k}\,\dddot{\bm k}+
    \frac32\ddot{\bm k}^2
    \right)+\cO(\bm k^2\,h)\,.
\ee
Therefore, we can view
the first line in \eqref{k exp} as the 
part of action free of
mutual interactions,
up to the $f$ dressing in $\bm \alpha$.
Note that the action for the spin-1 mode is negative-definite, and hence it has ``wrong'' overall sign. This leads to the instability of the colored JT gravity which was also observed in the 3D colored gravity~\cite{Gwak:2015jdo,Gwak:2015vfb,Joung:2017hsi}. We will discuss the consequence of this ``wrong'' sign in Section~\bref{sec: quadratic action}.
The second line in \eqref{k exp} describes
 the mutual interactions among the singlet graviton, the colored gravitons, and the spin-1 mode.

\subsection{Colored Isometry and Gauge Symmetry}
\label{sec: isometry of colored gravity}
%%%%%%%%%%%%%%%%%%%%%%%%%%%%%%%%%%%%%%%%%%%%%%%%%%%%%%%%%%%%%%%%%%%%%%%%%%%%

In Section~\bref{sec: isometry sl2} we have discussed the $SL(2,\mathbb R)$ redundancy
of $g$ in $a=g^{-1}\,\dot g$ \eqref{map_s}. In the color extended case, we find the same kind of redundancy: $a=g^{-1}\dot{g}=\tilde{g}^{-1}\dot{\tilde{g}}$, where  $\tilde g=k_0 g$ with a constant $k_0\in SU(N,N)$. Note that $k_0$ in the basis \eqref{SU(N,N)} is given by  
\be 
    k_0\,=\,\begin{pmatrix} 
    \bm D&\bm C\\
    \bm B&\bm A\\
    \end{pmatrix}:
\qquad
    k_0^\dag \begin{pmatrix}
    0 & - \bmI \\
    \bmI & 0\\
    \end{pmatrix} k_0 = \begin{pmatrix}
    0 & - \bmI \\
    \bmI & 0\\
    \end{pmatrix},
    \label{k elem}
\ee
that  can be rewritten as
\be 
    \bm A\,\bm C^\dagger=\bm A^\dagger\,\bm C\,,
    \qquad \bm B\,\bm D^\dagger=\bm D\,\bm B^\dagger\,,
    \qquad 
    \bm A\,\bm D^\dagger-\bm B^\dagger\,\bm C=\bm I\,. 
    \label{eq: su condition}
\ee
Similar to the $SL(2,\mathbb{R})$ case we need to mod out such a $SU(N,N)$ redundancy. To this end, using the decomposition~\eqref{g decomp} for  $g$ and $\tilde g$ related as $\tilde{g}=k_0\,g$ we find\footnote{These matrix transformations describe a color generalization of the modular transformations~$SL(2, \mathbb{Z})$, for example, of the elliptic genus.}
\begin{align}
    \tilde{\bm f}\,=\,&(\bm A\,\bm f+ \bm B)(\bm C\,\bm f+ \bm D)^{-1}\,,
    \label{f transf}\\
    \tilde{\bmd}^{-1}\, \tilde{\bmb}\,=\,& \bmd^{-1}\, \bmb \, (\bm C \, \bm f+ \bm D)^{-1}\,. \label{bd transf}
\end{align}
Let us note that one can find only the implicit expression for the transformation of $\bmb$ and $\bmd$. It is noteworthy that $\tilde{\bm f}$ is solely given by $\bm f$ through a form which generalizes the $SL(2,\mathbb R)$ transformation of the Schwarzian mode \eqref{eq: sl2 transformation}. 
A common constant factor in matrices $\bm A, \bm B, \bm C, \bm D$ --  which is the $\mathbb Z_{2N}$ phase $e^{i\,\frac{2\pi\,n}{2N}}$ -- does not affect the  transformation of $\bm f$~\eqref{f transf}, but  not those of $\bm b$ and $\bm d$. Thus,  the $\mathbb Z_{2N}$ phase factor
can be quotiented from  $\bm b$ and $\bm d$\,.

Recall that the classical solutions of the Schwarzian action are connected to one another by the isometry. Since one should mod out the isometry, it is enough to choose the simplest saddle classical solution $f(u)=u$. In the colored case, we shall also consider a series of classical solutions of the boundary effective action~\eqref{eq: boundary action zero temp}. Using the isometry~\eqref{f transf} and \eqref{bd transf} we can find a simplest solution which is connected to  others. 

Let us introduce  the following class of saddle classical solutions of the boundary action~\eqref{eq: boundary action zero temp}: 
\be 
    \bm f\,=\,\bm \cM\; u \,, 
    \qquad \bm d\,=\,{\rm constant}\,, 
    \hspace{8mm} \text{where}\qquad \bm\cM\,=\,{\rm constant}\,.\label{const sol}
\ee
Then, we consider transformations which map  solutions~\eqref{const sol} to each other. Using the $SU(N,N)$ defining conditions~\eqref{eq: su condition} we find that these are given by 
\be 
    \bm A\,=\,(\bm D^\dagger)^{-1}\,, 
    \quad \forall \bm D:\;\; \det \bm D \neq 0 
    \qquad \text{and} \quad
    \quad \bm B\,=\,\bm C\,=\,0\,.
    \label{const transf}
\ee

Then, consider the transformations \eqref{f transf} and \eqref{bd transf} in this particular case. By virtue of \eqref{const transf}  $\bm f$ transforms as 
\be 
    \tilde{\bm f}\,=\,\bm A\,\bm f\,\bm A^\dag\,, 
    \label{f D tr}
\ee 
that allows us to diagonalize Hermitian matrix $\bm f$ and, hence, $\dot{\bm f}$ by choosing $\bm A$ to be unitary. Indeed, due to the equation $\dot{\bm f}=\bm b^2$ \eqref{rels_col}, where $\bm b$ is Hermitian \eqref{g decomp}, the eigenvalues of Hermitian matrix $\dot{\bm f}$ should be positive  and can be normalized again using \eqref{f D tr}. In this way, we can set $\bm f=\bmI\,u$ so that its stabiliser group is $\bm A \in U(N)$. 

As for the transformation of  $\bm d$, by virtue of \eqref{bd transf} it is given implicitly through
\be 
\label{T_trans_su}
    \tilde{\bm b}\,\tilde{\bm d}
    \,=\,\bm A\,\bm b\,\bm d\,,
    \qquad 
    \tilde{\bm b}^{-1}\,\tilde{\bm d}
    \,=\,(\bm A^\dagger)^{-1}\, 
    \bm b^{-1}\,\bm d\,.
\ee
Restricting to the solution $\bm f=\bmI\,u$ (hence, $\bm b^2=\bmI$) and its stabiliser group $\bm A\in U(N)$, we find
\be 
    \tilde{\bm d}\,=\,\tilde{\bm A}\,\bm d\,,
\ee
where $\tilde{\bm A}=\tilde{\bm b}\,\bm A\,\bm b \in U(N)$. 
Since $\bm d\in SU(N)$, we can set $\bm d = \bm I$ by choosing an appropriate $\tilde{\bm A}$. With the remaining freedom $\bm A$ in $\tilde{\bm b}=\bm A\,\bm b=\bm b\,\bm A^\dagger$ \eqref{T_trans_su}, we can also fix $\bm b=\bmI$. 

Finally, we find that a class of solutions~\eqref{const sol}  defined  by the gauge function 
\be 
    g(u)\,=\,\begin{pmatrix}
    \bmI & \bm 0\\
    u\bmI&  \bmI
    \end{pmatrix}\,=\, \exp \big[ a_{\text{\tiny AdS}} u\big]\,,\hspace{8mm}   a_{\text{\tiny AdS}}\,=\, \begin{pmatrix}
    0 & 0 \\
    \bmI & 0\\
    \end{pmatrix} \,, \label{eq: classical sol zero t}
\ee
where $a_{\text{\tiny AdS}}$ corresponds to the AdS$_2$ background (set $\cL_0=0$ in \eqref{ads_sunn}).

%%%%%%%%%%%%%%%%%%%%%%%%%%%%%%%%%%%%%%%%%%%%%%%%%%%%%%%%%%%%%%%%%%%%%%
\subsection{Colored Schwarzian Action at Finite Temperature}
\label{sec: finite temperature}
%%%%%%%%%%%%%%%%%%%%%%%%%%%%%%%%%%%%%%%%%%%%%%%%%%%%%%%%%%%%%%%%%%%%%%

As we have discussed in Section~\bref{sec: decomposition}, the finite temperature JT gravity could be obtained by employing the Iwasawa decomposition which leads to a coordinate chart, where non-contractible cycle can be parameterized by a simple periodic boundary condition.
Note that  there is no simple expression  which relate the Iwasawa decomposition to the Gauss decomposition
for a general Lie group.\footnote{$G=KAN$, where $K$ is the maximal compact subgroup, $A$ and $N$ are the Abelian and the nilpotent subgroups, respectively.} In Section~\bref{sec:zero_temp color}, we have used a Gauss-like decomposition 
\eqref{g decomp} for  the zero temperature case. Therefore, in order to implement a finite temperature extension of the colored JT gravity we use 
an Iwasawa-like decomposition of $SU(N,N)$ which minimally modifies the decomposition \eqref{g decomp} to incorporate the maximal compact subgroup $K\subset SU(N,N)$.

It is known that $K \simeq  U(1)\otimes SU(N)\otimes SU(N)$ which can be parametrized as follows
\be
	K\,=\,\begin{pmatrix} 
	e^{i\,\theta}\,\bm d_1& \bm 0 \\ \bm 0 & e^{-i\,\theta}\,\bm d_2
	\end{pmatrix}
	\,=\,\begin{pmatrix} 
	\bm \xi & \bm 0 \\ \bm 0 & \bm \xi^{-1}
	\end{pmatrix}
	\begin{pmatrix} 
	\bm d& \bm 0 \\ \bm 0 & \bm d
	\end{pmatrix}. \label{eq: maximally compact subgroup}
\ee
Here, $\bm d_1, \bm d_2 \in SU(N)$ and $\theta\in \mathbb{R}$ parameterizes $U(1)$ factor, while $\bm \xi \in U(N)$ and $\bm d \in SU(N)$. The unitary transformation~\eqref{unitary} maps the matrix~\eqref{eq: maximally compact subgroup} to 
\be
\label{UYU}
	Y\,=\,U\,K\,U^{-1}
	\,=\,\begin{pmatrix}
	\frac{\bm \xi+\bm \xi^{-1}}2 & 
	\frac{\bm \xi-\bm \xi^{-1}}{-2\,i} \\
		\frac{\bm \xi-\bm \xi^{-1}}{2\,i} & 
	\frac{\bm \xi+\bm \xi^{-1}}2
	\end{pmatrix}
	\begin{pmatrix} 
	\bm d& \bm 0 \\ \bm 0 & \bm d
	\end{pmatrix}.
\ee
Hence, we introduce  the following Iwasawa-like decomposition of the group element $g(u)\in SU(N,N)$:
\begin{align}
	g(u)
	\,=\,&
	\begin{pmatrix}
	\frac{\bm \xi+\bm \xi^{-1}}2 & 
	\frac{\bm \xi-\bm \xi^{-1}}{-2\,i} \\
		\frac{\bm \xi-\bm \xi^{-1}}{2\,i} & 
	\frac{\bm \xi +\bm \xi^{-1}}2
	\end{pmatrix}
	\begin{pmatrix} 
	\bm b^{-1} & \bm 0 \\ \bm 0 & \bm b
	\end{pmatrix}
	\begin{pmatrix} 
	\bm d& \bm 0 \\ \bm 0 & \bm d
	\end{pmatrix}
	\begin{pmatrix} 
	\bm I & -\bm e \\ \bm 0 & \bm I
	\end{pmatrix}.
	\label{eq: iwasawa decomposition color}
\end{align}
As in Section~\bref{sec:zero_temp color}, going from $SU(N,N)$ to $PSU(N,N)$ we impose $\bm b\simeq -\bm b$ and $\bm d\in PSU(N)$.
This time, we find that
\be
	a(u)\,=\,g^{-1}\,\dot g
	\,=\,\begin{pmatrix}
	\bm p & \bm q \\
	\bm r & \bm s
	\end{pmatrix}\,, 
\ee
with 
\be
\begin{alignedat}{2}
	\bm p\,=\,&
	\bm d^{-1}\dot{\bm d}
	-\bm d^{-1}\,\dot{\bm b}\,\bm b^{-1}\,\bm d
	+\bm d^{-1}\,\bm b\,\bm\diffzeta \bm b^{-1}\,\bm d
	+\bm e\,\bm d^{-1}\,\bm b^{-1}\,
	\bm\theta\,\bm b^{-1}\,\bm d\,,
	 \\
\bm q\,=\,&-\dot{\bm e} -[ \bm d^{-1}\,\dot{\bm d},\bm e] + \bm d^{-1}\,\dot{\bm b}\,\bm b^{-1}\,\bm d\,\bm e
	 +\bm e\,\bm d^{-1}\,\bm b^{-1}\,\dot{\bm b}\,\bm d
	 \\ 
	 & 
	   +\bm e\,\bm d^{-1}\,\bm b^{-1}\,\bm \diffzeta\,\bm b\,\bm d
	  -\bm d^{-1}\,\bm b\,\bm \diffzeta\,\bm b^{-1}
	\,\bm d\,\bm e
	-\bm d^{-1}\,\bm b\,\bm \theta\,\bm b\,\bm d
	  -\bm e\,\bm d^{-1}\,\bm b^{-1}\,{\bm \theta}\,\bm b^{-1}\,\bm d\,\bm e\,, \\
	 \bm r\,=\,&
	  \bm d^{-1}\,\bm b^{-1}\,{\bm \theta}\,\bm b^{-1}\,\bm d\,, \\
	  & \bm s\,=\,  \bm d^{-1}\,\dot{\bm d} +\bm d^{-1}\,\bm b^{-1}\,\dot{\bm b}\,\bm d
    +\bm d^{-1}\,\bm b^{-1}\,\bm \diffzeta\,\bm b\,\bm d	   - \bm d^{-1}\,\bm b^{-1}\,{\bm \theta}\,\bm b^{-1}\,\bm d\,\bm e\,,\label{eq: component solution finite t color}
\end{alignedat}
\ee
cf. \eqref{g decomp}--\eqref{eq: sol a decomposition_21}. Here, $\bm \theta$ and $\bm \diffzeta$ are defined in terms of ${\bm \xi} \in U(N)$ as
\be
	\bm\theta\,=\,
	\frac{\bm \xi^{-1}\,\dot{\bm \xi}+\dot{\bm \xi}\,\bm \xi^{-1}}{2i}\,,
	\qquad
	\bm\diffzeta\,=\,	\frac{\bm \xi^{-1}\,\dot{\bm \xi}-\dot{\bm \xi}\,\bm \xi^{-1}}{2}\,.
	\label{th u}
\ee
The asymptotic AdS$_2$ condition~\eqref{cAAdS} gives
\begin{align}
& \bm\theta\,=\,\bm b^2\,,\label{eq: sol finite temp 1}\\
&  \bm d^{-1}\,\dot{\bm d} +\bm d^{-1}\,\bm b^{-1}\,{\partial_L} \bm b\,\bm d
 \,=\,i\,\bm\cJ+\bm e\,,\\
& 	\bm d^{-1}\dot{\bm d}
	-\bm d^{-1}\,\partial_R{\bm b}\,\bm b^{-1}\,\bm d
\,=\,i\,\bm\cJ-\bm e\,,\\
& \dot{\bm e}-\bm e^2+i\,[\bm \cJ,\bm e]+\bm d^{-1}\,\bm \theta^2\,\bm d\,=\,\bm \cL\,, \label{eq: sol finite temp 4}
\end{align}
where $\partial_L\bm b\,\equiv\,\dot {\bm b}+\bm\diffzeta\,\bm b$ and $\partial_R\bm b\,\equiv \,\dot {\bm b}-\bm b\,\bm\diffzeta$. From \eqref{eq: sol finite temp 1}--\eqref{eq: sol finite temp 4} we find the trace of $\bm \cL$ and $\bm \cJ^2$ as
\begin{align} 
	\tr(\bm\cL)\,=\,&\tr\left[\frac{d}{du}(\bm b^{-1}\,\dot{\bm b})
	- \left(\frac{\bm b^{-1}\partial_L{\bm b}+\partial_R{\bm b}\,\bm b^{-1}}2\right)^{\!\!2}
	-\bm b^4
	\right], \label{eq: matrix schwarzian lag at finite t}\\
	\tr(\bm \cJ^2)\,=\,&-\,\tr\left[(\bm d^{-1}\,\dot{\bm d})^2
	+\dot{\bm d}\,\bm d^{-1}\left(\bm b^{-1}\partial_L{\bm b}-\partial_R{\bm b}\,\bm b^{-1}\right)
	+\left(\frac{\bm b^{-1}\partial_L{\bm b}-\partial_R{\bm b}\,\bm b^{-1}}2\right)^2\right].
\end{align}
Combining these two terms we finally obtain the boundary effective action for the colored JT gravity at the finite temperature
\be
\begin{alignedat}{2} 
	S_{tot}\,=\,& - {\kappa \gamma \over N} \int \rd  u\; \tr\big( \bmcL+\bmcJ^2 \big)\\
	\,=\,&- {\kappa \gamma \over N} \int \rd  u \; \tr\left[\bm b^{-1}\ddot{\bm b}-2(\bm b^{-1}\dot{\bm b})^2
	-\bm b^4
	-\bm\diffzeta^2
	+\bm \diffzeta \left(\bm b^{-1}\dot{\bm b}-\dot{\bm b}\,\bm b^{-1}\right)\right.\\
	&\hspace{40mm} \left.-(\bm d^{-1}\,\dot{\bm d})^2
	-\dot{\bm d}\,\bm d^{-1}\left(\bm b^{-1}\partial_L{\bm b}-\partial_R{\bm b}\,\bm b^{-1}\right)
	\right].
	\label{f c Sch}
\end{alignedat}
\ee

Let us compare this result to that in the colored gravity at zero temperature \eqref{eq: boundary action zero temp} as well as in the uncolored  gravity at finite temperature  \eqref{f Sch}. First, note that the role of the variable $\bm f$ in the zero temperature case is now played by the variable $\bm \xi$. The former is $N\times N$ Hermitian matrix whereas the latter belongs to $U(N)$. In a sense, we  see a compactification of $N^2$ dimensional space by introducing a finite temperature. The equation  $\dot{\bm f}=\bm b^2$ \eqref{rels_col} is replaced by $\frac{\bm \xi^{-1}\,\dot{\bm \xi}+\dot{\bm \xi}\,\bm \xi^{-1}}{2i}=\bm b^2$ \eqref{th u}-\eqref{eq: sol finite temp 1}, and in the $N=1$ case the identification $\bm \xi=e^{i\,\theta/2}$ reproduces the equation \eqref{eq: sol sl2 finite temp} (up to rescaling) which leads to the (uncolored) Schwarzian action at finite temperature ~\eqref{f Sch}. The effective action \eqref{f c Sch} contains also the typical additional term $\bm b^4$ of the finite temperature Schwarzian which is analogous to the term ${1\over 2} \dot{\theta}^2$ in \eqref{f Sch}, but there are also a few other deviations from the zero temperature case, which are proportional to  $\bm\diffzeta$. 

If we re-introduce a variable $\bm f$ by parameterizing  $\bm \xi=\exp(i\bm f)$, then $\bm\theta$ and $\bm\diffzeta$  \eqref{th u} can be represented as\footnote{Note that the leading terms here are similar to the momentum and the $SU(N)$ angular momentum  of the free matrix quantum mechanics of $\bm f$.}  
\begin{equation}
    \bm\theta\,=\,\dot{\bm f}+\cO(\bmf^2)\,,
    \qquad 
    \bm\diffzeta\,=\,\frac12\,[\bmf,\dot{\bmf}]
    +\cO(\bmf^3)\, .
\end{equation}
This form implies that the terms proportional to $\bm\diffzeta$ in \eqref{f c Sch} can be viewed as self-couplings of the boundary colored spin-$2$ modes. Isolating the singlet graviton by virtue of the decomposition \eqref{singlet decmp} we see that it falls out of the leading contribution in $\diffzeta$. Nonetheless, it still contributes to the $\bm b^4$ term among additional terms in the finite temperature case.

The isometry which needs  to be modded  out in the boundary action at finite temperature can be obtained by repeating the same calculation as in  Section~\bref{sec: isometry of colored gravity} using the Iwasawa-like decomposition~\eqref{eq: iwasawa decomposition color}. We find that
\be
\label{eq: colored isometry 1 finite t}
\tnfunc\, (\tilde{\bm \xi}) \; \,=\, \big(\bm A\,  \tnfunc\,(\bm \xi)+ \bm B\big)\big(\bm C\,  \tnfunc\,(\bm \xi)+ \bm D\big)^{-1},
\ee
\be
\label{eq: colored isometry 2 finite t}
\tilde{\bm d}^{-1}\, \tilde{\bm b} \; \scfunc\,(\tilde{\bm \xi}) \,=\,  \bm d^{-1}\,\bm b \; \scfunc\,(\bm \xi) \, \bigg[  \bm C\,  \tnfunc\,(\bm \xi) +\bm D \bigg]^{-1}, 
\ee
where the tangent-like $\tnfunc(\bm \xi)$ and secant-like function $\scfunc(\bm \xi)$ of the matrix $\bm \xi$ are defined by
\be
\tnfunc\, (\bm \xi)\,\equiv \, {1\over 2i } \big(\bm \xi - \bm \xi^{-1}\big)\bigg[ {1\over 2}\big(\bm \xi + \bm \xi^{-1} \big) \bigg]^{-1},
\qquad
\scfunc\,(\bm \xi)\,\equiv \,  \bigg[{1\over 2} \big(\bm\xi'+\bm\xi'^{-1}\big)\bigg]^{-1}.
\ee
The isometry~\eqref{eq: colored isometry 1 finite t},  \eqref{eq: colored isometry 2 finite t} generalizes the respective transformations \eqref{T_trans_sl2} in the $SL(2, \mathbb{R})$ finite temperature case as well as transformations \eqref{f transf} and \eqref{bd transf} in the $SU(N,N)$ zero temperature case.
%
%\begin{align}
%    \text{tg}\, (\bm \xi)\,\equiv \,& {1\over 2i } \big(\bm \xi - \bm \xi^{-1}\big)\bigg[ {1\over 2}\big(\bm \xi + \bm \xi^{-1} \big) \bigg]^{-1} \,,\\
%    \text{sc}\,(\bm \xi)\,\equiv \, & \bigg[{1\over 2} \big(\bm\xi'+\bm\xi'^{-1}\big)\bigg]^{-1}\
%\end{align}

%%%%%%%%%%%%%%%%%%%%%%%%%%%%%%%%%%%%%%%%%%%%%%%%%%%%%%%%%%%%%%%%%%%%%%%%%%%%%%
\subsection{Holonomy for $SU(N,N)$ Connection}
\label{sec: smooth background}
%%%%%%%%%%%%%%%%%%%%%%%%%%%%%%%%%%%%%%%%%%%%%%%%%%%%%%%%%%%%%%%%%%%%%%%%%%%%%%

In Section~\bref{sec: isometry of colored gravity} we have considered a class of classical solutions of the boundary effective action which  is related by the isometry to the  AdS$_2$ background at zero temperature \eqref{eq: classical sol zero t}. At finite temperature, we can also consider the following simple classical solutions of the boundary effective action~\eqref{f c Sch}:
\begin{align}
    \bm b\,=\,\bm b_0\,, \qquad \bm \theta \,=\,\bm b_0^2   \,, \qquad   \bm d\,=\, \bmI \,, \qquad \diffzeta\,=\, \bm e\, =\,\bm 0\,, 
\end{align}
where $\bm b_0$ is a constant $N\times N$ Hermitian matrix. Using the relations \eqref{eq: component solution finite t color} one can see that this solution corresponds to the constant connection  $a_0$ given by
\begin{align}
    a_0\,=\, g^{-1}\dot{g}\,=\, \begin{pmatrix}
    \bm 0 & - \bmcL_0\\
    \bmI & \bm 0\\
    \end{pmatrix}, \qquad \mbox{where} \qquad \bmcL_0\,\equiv \,\bm b_0^4\,. 
    \label{eq: constant gauge field a color finite t}
\end{align}
However, for a given temperature $T=\beta^{-1}=(2\pi)^{-1}$, not all the constant $\bmcL_0$ 
can be obtained from a single-valued gauge function
as in Section~\bref{sec: asympototic ads},
and this issue is captured by 
 the holonomy of the gauge field $A$ along the thermal circle,
\begin{align}
    \text{Hol}(A)\,=\, \cP \exp \bigg[ \oint A\bigg] \,\sim\, \exp \big(2\pi a_0\big)\,,
\end{align}
where $A=b^{-1}({\rm d} +a) b$, and $a$ is a fluctuation around the constant background $a_0$~\eqref{eq: constant gauge field a color finite t}. Like in the $SL(2,\mathbb{R})$ case in Section~\bref{sec: isometry of colored gravity}, one can assume that a ``nice geometry'' in the colored gravity
have a single-valued gauge function,
and hence a trivial holonomy. 
Viewed as an $SU(N,N)$ element,
 the trivial holonomy belongs to the center subgroup $\mathbb Z_{2N}$ of $SU(N,N)$:
\begin{align}
    \text{Hol}(A)\,\sim\, e^{2\pi a_0}\,=\, e^{{\pi i n \over N}} I_{2N}\,\in\, Z\big[SU(N,N)\big] \hspace{8mm}(n=0,1,\cdots, 2N-1)\label{eq: smoothness condition}\,. 
\end{align}
Using a gauge transformation by a constant gauge parameter
\begin{align}
    \begin{pmatrix}
    \bmU^{-1} & 0\\
    0 & \bmU^{-1}\\
    \end{pmatrix}a_0\begin{pmatrix}
    \bmU & 0\\
    0 & \bmU\\
    \end{pmatrix}\,=\, \begin{pmatrix}
    0 & -\bmU^{-1}\bmcL_0\bmU\\
    \bmI & 0\\
    \end{pmatrix},\quad\text{where}\quad \bmU\in U(N):\;\;\mbox{constant}\,, 
\end{align}
one can diagonalize the matrix $\bmcL_0$ as
\begin{align}
    \bmcL_0\,=\, \text{diag}\big(\lambda_1,\lambda_2,\cdots,\lambda_N\big)\,,
\end{align}
with eigenvalues $0\leqq \lambda_1\leqq \lambda_2\leqq \cdots\leqq \lambda_N$. Note that the eigenvalues of $\bmcL_0$ are non-negative because $\bmcL_0=\bm b_0^4$ with Hermitian matrix $\bm b_0$  \eqref{eq: constant gauge field a color finite t}. Then, the $2N$ eigenvalues of the constant $a_0$ are found to be
\begin{align}
\label{a_eigen}
    - i \,\sqrt{\lambda_1}\,,\;- i \,\sqrt{\lambda_2}\,,\;\cdots\,,\;- i \,\sqrt{\lambda_N}\,,\;  i \,\sqrt{\lambda_1}\,,\; i \,\sqrt{\lambda_2}\,,\;\cdots\,,\; i \,\sqrt{\lambda_N}\,. 
\end{align}
Therefore, from the condition~\eqref{eq: smoothness condition}, we conclude that the holonomy $\text{Hol}(A)$  can be $\pm I_{2N}$, and the eigenvalues of the corresponding $\bmcL_0$ are given by
\begin{align}
\label{eigen1}
    \lambda_j\,=\, %{\pi^2\nu_j^2 \over \beta^2}
    {\nu_j^2 \over 4}\;,\hspace{8mm} \mbox{where}\quad \nu_j\in \mathbb{Z}_+\,\quad\text{and}\quad \nu_1\leqq \cdots\leqq \nu_N\,.
\end{align}
Here, all $\nu_j$'s are either even for $\text{Hol}(A)\,=\,I_{2N}$ or odd for $\text{Hol}(A)\,=\,-I_{2N}$.

The Hamiltonian of the Schwarzian theory is defined by the Schwarzian derivative up to a factor~\cite{Stanford:2017thb} and one can rewrite the Hamiltonian density in terms of the variable $a(u)$ \eqref{eq: gauge field sol sl2 finite temp} as
\begin{align}
    \cH\,=\, -{\kappa \gamma\over 2}\, \bigg(\text{Sch}\big[f(u),u\big]+{1\over 2} \dot{f}^2(u)\bigg)\,=\,{\kappa \gamma\over 2} \,\tr (a^2)\,.
\end{align}
Using this expression  we define the energy of the colored gravity by
\begin{align}
\label{energy_def}
    E\,\equiv\, {\kappa \gamma \over 2 N} \int^\beta_0 \rd  u\;\tr (a^2)\,.
\end{align}
Especially, the energy of the constant solution $a_0$ is 
\be
E_0\,=\, {\beta \kappa \gamma\over 2N} \, \tr a_0^2\,=\,-{\pi^2\kappa \gamma\over \beta N }\sum_{j=1}^N \nu_j^2\,, \label{eq: energy of constant background}
\ee
where we retrieved the temperature $T=\beta^{-1}$. The highest-energy smooth constant solution has $\nu_j=1$ for all $j=1,2,\cdots, N$, and this is the  global AdS$_2$ background with $\bmcL_0={\pi^2\over \beta^2} \bmI$ (see discussion in Section \bref{sec:back_sol}). The other constant solutions have lower energy.
These are analogous to the conical surplus in the 3D Chern-Simons (higher-spin) gravity~\cite{Castro:2011iw}, which is often considered as ``unphysical'' one. We also observed the signal of its unphysical nature in Section~\bref{sec: quadratic action}.

%%%%%%%%%%%%%%%%%%%%%%%%%%%%%%%%%%%%%%%%%%%%%%%%%%%%%%%%%%%
\subsection{Asymptotic AdS$_2$ Symmetry}
\label{sec: asymptotic symmetry}
%%%%%%%%%%%%%%%%%%%%%%%%%%%%%%%%%%%%%%%%%%%%%%%%%%%%%%%%%%%

In Section~\bref{sec:zero_temp color} we have imposed the asymptotic AdS$_2$ condition and the gauge condition for $a(u)$  as follows (see \eqref{as_1}, \eqref{cAAdS})
\begin{align}
\label{asym_cond}
    a(u)\,=\,L_1\otimes \bmI  + i I \otimes \bmcJ(u)  + L_{-1} \otimes \bmcL(u)   \,=\, \begin{pmatrix}
    i \bmcJ(u)  & - \bmcL(u) \\
    \bmI  & i \bmcJ(u)  \\
    \end{pmatrix}\,.
\end{align}
In this section, we will consider the residual gauge symmetry which keeps the form of $a(u)$ \eqref{asym_cond} intact. This will lead to the asymptotic AdS$_2$ symmetry for the colored JT gravity. Using \eqref{X_par} the gauge parameter $h(u) \in su(N,N)$ can be represented as
\begin{align}
    h(u)\,=\,\begin{pmatrix}
    \bml & \bmn \\
    \bmm & -\bml^\dag \\
    \end{pmatrix}, \quad \text{where}\;\; \bml\,\equiv\,\bm s+i \bm t\,.
\end{align}
where $\bmm$, $\bmn$, $\bms$, $\bmt$ are Hermitian $N\times N$ matrices. Then,  an infinitesimal gauge transformation of $a(u)$ can be written as
\be
\delta a(u) = \dot{h} + [a(u),h]\,=\begin{pmatrix}
    \dot{\bml} & \dot{\bmn} \\
    \dot{\bmm} & -\dot{\bml}^\dag  \\
    \end{pmatrix} + 
    \begin{pmatrix}
    -\bmn -  \bmcL \,\bmm +i [\bmcJ,\bml ] \quad\ & \bmcL\,\bml^\dag + \bml\,\bmcL + i [\bmcJ,\bmn] \\
    2\bms+ i [\bmcJ,\bmm] & \bmn+\bmm\,\bmcL -i[\bmcJ,\bml^\dag]  \\
    \end{pmatrix}\,.\label{eq: inf residual gauge transf}
\ee
Imposing the condition $\delta a(u)=0$ we find the following constraints on  the gauge parameters 
\be
\label{eq: asym ads sol color 2}
\bms\,=\,-\half  \cD_{\cJ} \bmm\,, 
\qquad
\bmn \,=\,\cD_{\cJ} \bms - \half \{\bmcL, \bmm\}\,, 
\ee
where $\cD_{\cJ}\,\equiv\, \partial_u + i[\bmcJ, \;\;]$. Hence,  the residual transformations can be parameterized by $\bmm$ and $\bmt$, and the corresponding transformations of $\bmcJ$ and $\bmcL$ are found to be
\be
\label{eq: color asymptotic ads transf 1}
\delta \bmcJ =  \cD_{\cJ} \bmt + {i\over 2}\,[\bmcL,\bmm]\,, 
\qquad
\delta \bmcL =  \half\cD_{\cJ}^3 \bmm + \half \{ \cD_{\cJ} \bmcL ,\bmm\} +\{\bmcL, \cD_{\cJ} \bmm\}+i\, [\bmcL, \bmt]\,. 
\ee
To identify transformations of each component, we decompose $\bmcL$ and $\bmcJ$ as well as the gauge parameters $\bmm$ and $\bmt$ in the $u(N)$ basis as follows 
\begin{align}
    \bmcL\,=\,& \cL \,\bmI + \bmcK\,=\, \cL \,\bmI + \cK^A \, \bT_A\,, &\bmm\,=\,&\xi\, \bmI + \bmzeta\,=\, \xi \,\bmI + \zeta^A \,\bT_A \,, \\
    \bmcJ\,=\,& \cJ^A\, \bT_A\,, & \bmt\,=\,& \bmlambda\,=\, \lambda^A\, \bT_A\;,
\end{align}
(see the beginning of Section \bref{sec: color generalization}). The transformation of $\cJ^A$ is found to be
\begin{align}
\label{tra1}
    \delta \cJ^A\,=\,  \dot{\lambda}^A -2 {f_{BC}}^A \cJ^B \lambda^C+ {f_{BC}}^A \cK^B \zeta^C\,,
\end{align}
where $f_{ABC}$ are the $su(N)$ structure constants. In addition, the $\cL$ is transformed as
\be
\label{tra2}
\ba{c}
\dps
    \delta \cL \,=\, {1\over N}\tr \big[\delta \bmcL\big] 
    \,=\, -{1\over 2}\dddot \xi+ \dot{\cL}\, \xi + 2\cL \,\dot{\xi} + {1\over N} (\cD_{\cJ}\cK)^A\zeta_A + {2\over N}(\cD_{\cJ} \zeta)^A \cK_A\,, 
\vspace{2mm}
\\
\dps
\,=\, {1\over 2}\dddot \xi+ \dot{\cL}\, \xi + 2\cL\, \dot{\xi} + {1\over N} \dot{\cK}^A\,\zeta_A + {2\over N}\cK^A\,\dot{\zeta}_A + {2\over N} {f_{ABC}} \cJ^A\, \cK^B \,\zeta^C\,,
\ea
\ee
where we used the product relation \eqref{prod_T}.
One can also obtain the transformation 
\begin{align}
\label{tra3}
    \delta \cK^A \,=\,\tr\big[ \delta\bmcL\, \bmT^A \big]\,=\, {1\over 2}\partial_u^3\zeta^A+  \dot{\cL} \,\zeta^A + 2 \cL\, \dot{\zeta}^A +  \dot{\cK}\, \zeta^A + 2 \cK\, \dot{\zeta}^A + \cdots\,.
\end{align}
The transformations \eqref{tra1}-\eqref{tra3} are analogous to the asymptotic colored AdS$_2$ symmetry of the 3D colored Chern-Simons gravity~\cite{Joung:2017hsi}.

%%%%%%%%%%%%%%%%%%%%%%%%%%%%%%%%%%%%%%%%%%%%%%%%%%%%%%%%%%%%%%%%%%%%%%%
\section{Quantum Fluctuations around Classical Solutions}
\label{sec: quadratic action}
%%%%%%%%%%%%%%%%%%%%%%%%%%%%%%%%%%%%%%%%%%%%%%%%%%%%%%%%%%%%%%%%%%%%%%%

In Section~\bref{sec: finite temperature} we have obtained the boundary effective action at  finite temperature. However, it was difficult to express the action in terms of $\bm \xi$ which is analogous to $\bmf$ in the zero-temperature case, and, therefore, a perturbative expansion of the boundary action around the finite temperature background would not be straightforward as in the $sl(2,\mathbb{R})$ Schwarzian theory. A similar difficulty also appears in the higher-spin gravity where the closed form of the boundary effective action at finite temperature is still an open question. Nevertheless, one can work out  a perturbative analysis for the case of higher-spin gravity at finite temperature \cite{Narayan:2019ove}. For example, in the 3D Chern-Simons (higher-spin) gravity, starting from some background connection, one can perturbatively build an infinitesimal gauge transformation which keeps the asymptotic AdS condition intact. It maps the background connection to some new gauge connection parameterized by the gauge parameters which therefore describe the physical boundary modes. Then, provided the appropriate boundary condition, one can evaluate the total action on the resulting gauge connection. This leads to the boundary action  expressed perturbatively in terms the boundary modes  propagating on the fixed background (e.g. BTZ black hole). In the same way, we will also consider an infinitesimal gauge transformation~\eqref{eq: inf residual gauge transf} around the  background solution found in Section~\bref{sec: smooth background} to derive perturbatively the quadratic boundary action of the colored JT gravity.

\subsection{Boundary Effective Action}

Recall that the finite temperature boundary action~\eqref{f c Sch}  was obtained by using the representation $a(u)=g^{-1} \dot{g}$ \eqref{map_s}. As  explained in Section~\bref{sec: isometry sl2}, the connection $a(u)$ can also be  parameterized by a smooth gauge transformation of the constant connection $a_0$~\eqref{eq: constant gauge field a color finite t}\footnote{In general, one may consider a background with non-zero $\bmcJ_0$ in $a_0=\begin{pmatrix} i \bmcJ_0 & -\bmcL_0\\
\bmI & i\bmcJ_0 \\
\end{pmatrix}$. For simplicity, we analyze the background with $\bmcJ_0\,=\,0$.}
\begin{align}
    a_0\,=\,\begin{pmatrix}
    \bm 0 & -\bmcL_0\\
    \bmI & \bm 0\\
    \end{pmatrix}\,.  \label{eq: color constant gauge field}
\end{align}
Therefore, an infinitesimal gauge transformation which keeps the asymptotic AdS$_2$ condition~\eqref{asym_cond} intact can produce the perturbation of the boundary action~\eqref{f c Sch}. To this end, let us consider a smooth gauge transformation of the constant connection  $a_0$ by the gauge group element $U(u)\in SU(N,N)$ which is expanded with respect to a small parameter $\epsilon$ as follows
\be
\label{gauge_p_exp}
U(u)\, =\, I_{2N} \, +\, \epsilon \, h(u) \,+\, {1\over 2} \epsilon^2 \,\big( \,[h(u)]^2 +k(u)\, \big)\,+\, \mathcal{O}(\epsilon^3)\,,
\ee
where $I_{2N}$ is $2N\times 2N$ unit  matrix.  From the condition~\eqref{SU(N,N)} for a $SU(N,N)$ element, we have 
\begin{align}
\label{gauge_p_hk}
    h(u)=&\begin{pmatrix}
    \bms(u)+ i \bmt(u) & \bmn(u)\\
    \bmm(u) & -\bms+i \bmt(u)\\
    \end{pmatrix},
    \qquad 
    k(u)=\begin{pmatrix}
    \bm \sigma(u)+ i \bm \tau(u) & \bm \nu(u)\\
    \bm \mu(u) & -\bm \sigma(u)+i \bm \tau(u)\\
    \end{pmatrix}. 
\end{align}
Here, matrices $\bmm$, $\bmn$, $\bms$, $\bmt$ and  $\bm \mu$, $\bm \nu$, $\bm \sigma$, $\bm \tau$ are Hermitian. The perturbative expansion of the smooth gauge transformation  is found to be
\be
\label{eq: expansion of a color}
\begin{alignedat}{2}
a(u)  \,=\, & U^{-1} a_0 U + U^{-1}\dot U
\vspace{2mm}
\\
\,=\,&a_0 + \epsilon \left(\dot{h} +[a_0,h] \right)+\half\epsilon^2\left(\dot{k} + [a_0,k] +\dot{h}\,h-h\,\dot{h}+a_0h^2-2ha_0h + h^2a_0\right)\\
&+\cO(\epsilon^3)\,.
\end{alignedat}
\ee
At order $\mathcal{O}(\epsilon)$, the asymptotic AdS$_2$ condition and the gauge condition~\eqref{cAAdS} allow to express $\bms$ and $\bmn$ in terms of $\bmm$ as in Section~\bref{sec: asymptotic symmetry},
\be
\label{eq: gauge parameter aads sol 1}
    \bms=-\frac12\dot{\bmm}\,,
    \qquad
    \bmn= -\frac12\ddot{\bmm} -\frac12 \{ \bmcL_0, \bmm\}\,. 
\ee
In the same way, the condition~\eqref{cAAdS} at order $\mathcal{O}(\epsilon^2)$ gives
\begin{align}
    \bm \sigma\,=\,& -{1\over 2}\dot{\bm\mu} + {i\over 2} [\bm\mu,\dot{\bmt}]
    +{1\over 4} \{\bmcL_0,\bmm^2\}- {1\over 2}\bmm\,\bmcL_0\, \bmm\,,  \label{eq: gauge parameter aads sol 3}\\
    \bm\nu\,=\,&-{1\over 2}\, \ddot{\bm\mu} -{1\over 2} \{\bmcL_0, \bm\mu\}+{1\over 4} \{\bmcL_0, \bmm^2\}- {1\over 2}\bmm\,\bmcL_0\, \bmm \cr
    &-{1\over 4}\,\{ \bmm,\dddot{\bmm}\} +{3\over 4}\, [[\bmcL_0,\bmm],\dot{\bmm}]-{1\over 2}\,\{\{\bmm,\dot{\bmm}\},\bmcL_0\}\cr
    &+i\, [\dot{\bmm},\dot{\bmt}]+{i\over 2}\, [\dot{\bmm},\ddot{\bmt}]+ {i\over 2}\, \left\{ \bmm,[\bmt,\bmcL_0]\right\}\,. \label{eq: gauge parameter aads sol 4}
\end{align}
Expanding the connection $a$ in \eqref{eq: expansion of a color} yields  the perturbative expansion of $\bmcL$ and $\bmcJ$ with respect to $\epsilon$
\be
\bmcL\,=\, \bmcL_0 + \epsilon \bmcL^{(1)}+\epsilon^2 \bmcL^{(2)}+\cdots\,,
\qquad 
\bmcJ\,=\, \epsilon \bmcJ^{(1)}+\epsilon^2 \bmcJ^{(2)}+\cdots\,.
\ee
Similarly, the action can be expanded  as follows  
\begin{equation}
\begin{alignedat}{2}
\label{eq: bdy action rainbow3}
\dps
S_{tot}\,=\,& - {\kappa \gamma \over N} \int \rd  u\; \tr\big( \bmcL+\bmcJ^2 \big)  \\ 
\,=\, &
-{\kappa \gamma \over  N} \int \rd u\;\tr 
\left[ \bmcL_0 -\epsilon\,  \bmcL^{(1)}
-\epsilon^2 \left(
\bmcL^{(2)}+\big(\bmcJ^{(1)}\big)^2\right) + \cO(\epsilon^3)
\right]\,.
\end{alignedat}
\end{equation}
The leading order term is the energy \eqref{energy_def} of the constant background connection  $a_0$ given by \eqref{eq: color constant gauge field}. Using the solution~\eqref{eq: gauge parameter aads sol 1}--\eqref{eq: gauge parameter aads sol 4} for the asymptotic AdS$_2$ condition, we can express $\bmcL^{(n)}$ and $\bmcJ^{(n)}$ ($n=1,2$) in terms of $\bmm$ and $\bmt$. Especially, we are interested in $\tr(\bmcL^{(1)})$, $\tr(\bmcL^{(2)})$ and $\tr\bigg[\big(\bmcJ^{(1)}\big)^2\bigg]$ in \eqref{eq: bdy action rainbow3}. Up to total derivatives we find
\begin{align}
    \int \rd u\;\tr(\bmcL^{(1)})\,=\,&0\,, \label{trL1} \\
    \int \rd u\;\tr(\bmcL^{(2)})\,=\,&\int du\;\tr\bigg( -{1\over 4} \, \ddot{\bmm}^2 +\bmcL_0 \dot{\bmm}^2 + {1\over 2}\left(\bmcL_0 \bmm\right)^2 - {1\over 2} \bmcL_0^2 \bmm^2  \bigg)\,, \\
    \int \rd u\; \tr\big(\bmcJ^{(1)}\big)^2\,=\,&\int \rd u\; \tr \bigg(\dot{\bmt}^2 +i\dot{\bmt}[\bmcL_0,\bmm] - {1\over 2}\left(\bmcL_0 \bmm\right)^2 + {1\over 2} \bmcL_0^2 \bmm^2 \bigg)\,. 
\end{align}
In total, the perturbative expansion of the boundary action~\eqref{f c Sch} up to quadratic order is found to be
\be
\label{eq: boundary action expansion color}
\begin{alignedat}{2}
    S_{tot}\,=\,& -{\kappa \gamma\over N} \int \rd u \; \tr \big(\bmcL+\bmcJ^2\big)\\
    \,=\,&E_0 + {\kappa \gamma \epsilon^2 \over N} \int \rd u \; \tr \bigg( {1\over 4} \, \ddot{\bmm}^2 -\bmcL_0 \dot{\bmm}^2  - \dot{\bmt}^2 - i\dot{\bmt}[\bmcL_0,\bmm] \bigg)+ \mathcal{O}\big(\epsilon^3\big)\,, 
\end{alignedat}
\ee
where the energy $E_0$ is given by \eqref{eq: energy of constant background}. Note that in the  AdS$_2$ background, $\bmcL_0={\pi^2 \over\beta^2}\bmI$, the singlet graviton, colored graviton and spin-$1$ modes are decoupled at quadratic level. Moreover,  the spin-$1$ mode $\bmt$ has a wrong-sign kinetic term that leads to instability of the colored JT gravity. This instability persists even for the other constant backgrounds $\bmcL_0$ (see the next section).

\subsection{Mode Expansion}
\label{sec:modes}

As is shown in \eqref{eigen1}, the smooth background $\bmcL_0$ can be chosen to be
\begin{align}
    \bmcL_0= {\pi^2 \over \beta^2 }\,\text{diag}\big(\,\nu_1^2,\nu_2^2 ,\cdots, \nu_N^2\, \big) \,, \hspace{4mm}\text{where}\quad \nu_j\in\mathbb{Z}_+ \quad\text{and} \quad \nu_1\leqq \nu_2\leqq \cdots \leqq \nu_N\,, \label{eigen2}
\end{align}
where we retrieved the temperature $\beta^{-1}$. Then, in terms of the matrix entries 
\be
f_{jk}\equiv(\bmm)_{jk}
\quad 
\text{and}
\quad
\phi_{jk}\equiv(\bmt)_{jk}\;, \qquad j,k=1,2,\cdots, N\,,
\ee
the quadratic part $S_{tot}^{(2)}$ of the boundary action~\eqref{eq: boundary action expansion color} can be written as
\be
\begin{alignedat}{2}
    S^{(2)}_{tot}%\,=\,  {\kappa \gamma \over N} \int du \; \sum_{j,k}\bigg[{1\over 4}\ddot{f}_{jk}\ddot{f}_{kj}-{\pi^2 \nu_i^2\over \beta^2 } \dot{f}_{jk}\dot{f}_{kj}- \dot{\phi}_{jk} \dot{\phi}_{kj} + {i\pi^2\over \beta^2} (\nu_j^2-\nu_k^2) \dot{\phi}_{jk} f_{kj} \bigg]\cr
    =&{\kappa \gamma \over N} \int \rd u \; \sum_{j=1}^N\bigg[{1\over 4}\ddot{f}_{jj}\ddot{f}_{jj}-{\pi^2 \nu_j^2\over \beta^2 } \dot{f}_{jj}\dot{f}_{jj}- \dot{\phi}_{jj} \dot{\phi}_{jj} \bigg]\cr
    &+{2\kappa \gamma \over N} \int \rd u \; \sum_{j>k}\bigg[{1\over 4}\ddot{f}_{jk}\ddot{f}_{jk}^\ast-{\pi^2 (\nu_j^2+\nu_k^2)\over 2\beta^2 } \dot{f}_{jk}\dot{f}_{jk}^\ast- \dot{\phi}_{jk} \dot{\phi}_{jk}^\ast  \bigg]\cr
    &+{2\kappa \gamma \over N} \int \rd u \; \sum_{j>k}\bigg[ {i\pi^2\over \beta^2} (\nu_j^2-\nu_k^2) (\dot{\phi}_{jk} f_{jk}^\ast- \dot{\phi}_{jk}^\ast f_{jk}) \bigg]\,. 
\end{alignedat}
\ee
Since the temperature is finite (i.e. $u\in S^1$), we can use the Fourier mode expansion of $f_{jk}(u)$ and $\phi_{jk}(u)$,
\begin{align}
    f_{jk}(u)\,=\,& {1\over \sqrt{2\pi}}\sum_{n\in \mathbb{Z}} f_{jk,n} \, e^{-{2\pi i n u\over \beta}}\,, \\
    \phi_{jk}(u)\,=\, & {1\over \sqrt{2\pi}} {2\pi \over \beta}  \sum_{n\in \mathbb{Z}} \phi_{jk,n}\,e^{-{2\pi i n u\over \beta}}\,, 
\end{align}
so that the quadratic boundary action can be expanded as
\be
\label{eq: quadratic action colored gravity}
\begin{alignedat}{2}
     S_{tot}^{(2)}\,=\,&{\kappa \gamma\over  N} \bigg({2\pi \over \beta}\bigg)^4\bigg[{1\over 4}\sum_{j=1}^N \sum_{n} n^2\big(n^2 - \nu_j^2 \big)\, f_{jj,-n}f_{jj,n} - \sum_{j=1}^N \sum_{n\in \mathbb{Z}} n^2\, \phi_{jj,-n}\phi_{jj,n}\bigg]\cr
%     &+ {\kappa \gamma \over 2N }  \bigg({2\pi \over \beta}\bigg)^4\sum_{j>k} n^2\bigg( n^2 -{\nu_j^2+\nu_k^2\over 2}\bigg)\bar{f}_{jk,n} f_{jk,n}- {2\kappa \gamma \over N }  \bigg({2\pi \over \beta}\bigg)^4\sum_{j>k} n^2\, \bar{\phi}_{jk,n}\phi_{jk,n}\cr
%     & + {\kappa \gamma\over 2N} \bigg({2\pi  \over \beta}\bigg)^4 \sum_{j>k} (\nu_j^2-\nu_k^2)n(\phi_{jk,n}\bar{f}_{jk,n}+ \bar{\phi}_{jk,n}f_{jk,n} )
    &+ {\kappa \gamma \over 2N }  \bigg({2\pi \over \beta}\bigg)^4\sum_{j>k}\sum_{n\in \mathbb{Z}}  \begin{pmatrix}
    f^\ast_{jk,n} & \phi^\ast_{jk,n}\\
    \end{pmatrix}\cK_{n}^{(j,k)} \begin{pmatrix}
    f_{jk,n}\\
    \phi_{jk,n}\\
    \end{pmatrix}\,, 
\end{alignedat}
\ee
where the $2\times 2$ matrix $\cK^{(j,k)}_{n}$ is given by
\begin{align}
    \cK_{n}^{(j,k)}\,\equiv\, \begin{pmatrix}
    n^2\bigg[ n^2 -{\nu_j^2+\nu_k^2\over 2}\bigg] & (\nu_j^2-\nu_k^2)n\\
    (\nu_j^2-\nu_k^2)n & -  n^2\\
    \end{pmatrix}. 
\end{align}
In order to diagonalize the quadratic action~\eqref{eq: quadratic action colored gravity} we find the eigenvalues $\lambda_{n,\pm}^{(j,k)}$ of the matrix $\cK_{n}^{(j,k)}$:
\begin{align}
    \lambda_{n,\pm}^{(j,k)}\,=\,{n\over 16}\bigg[2n^3 - n\big(2 + \nu_j^2+\nu_k^2\big)\pm \sqrt{n^2\big(\nu_j^2+\nu_k^2-2n^2-2\big)^2+16\big(\nu_j^2-\nu_k^2\big)^2 } \bigg]\,. 
\end{align}
Introducing suitable eigenfunctions $\xi_{jk,n;\pm}$ the quadratic action $S^{(2)}_{tot}$ can be cast into the form 
\be
\label{eq: color quadratic action diagonal}
\begin{alignedat}{2}
     S_{tot}^{(2)}\,=\,&{\kappa \gamma\over  N} \bigg({2\pi \over \beta}\bigg)^4\bigg[{1\over 4}\sum_{j=1}^N \sum_{n\in \mathbb{Z}} n^2\big(n^2 - \nu_j^2 \big)\, f_{jj,-n}f_{jj,n} - \sum_{j=1}^N \sum_{n\in \mathbb{Z}} n^2\, \phi_{jj,-n}\phi_{jj,n}\bigg]\cr
    &+ {\kappa \gamma \over 2N }  \bigg({2\pi \over \beta}\bigg)^4\sum_{j>k}\sum_{n\in \mathbb{Z}} \bigg[\lambda^{(j,k)}_{n,+} \xi_{jk,n;+}^\ast\, \xi_{jk,n;+}+\lambda^{(j,k)}_{n,-}\, \xi_{jk,n;-}^\ast \xi_{jk,n;-}\bigg]\,. 
\end{alignedat}
\ee
Note that eigenvalues $\lambda_{n,-}^{(j,k)}$ become negative for sufficiently  large $n$ that  leads to the instability of the colored gravity. From the quadratic action~\eqref{eq: color quadratic action diagonal} one can read off the 2-point functions: 
\be
\label{eq: color propagator 1}
\langle f_{jj,-n} \,f_{jj,n} \rangle \sim {1\over n^2(n^2-\nu_j^2)}\,,
\quad
\langle \phi_{jj,-n}\, \phi_{jj,n} \rangle\sim {1\over n^2}\,, 
\quad
\langle \xi_{jk,n;\pm}^\ast \, \xi_{jk,n;\pm} \rangle\sim{1\over \lambda^{(j,k)}_{n,\pm}}\,. 
\ee
Note that some modes, e.g. $f_{jj,n}$ at $n=\nu_j$, vanish in the quadratic action, which leads to the divergence of the propagator. However, those modes correspond to the isometry of the constant background $a_0$~\eqref{eq: color constant gauge field},
which we regard as  ``gauge symmetries''.
Therefore,
we exclude their contributions in the path-integral.  In details, it is easy to see that $f_{jj,0}$, $f_{jj,\nu_j}$, and $\phi_{jj,0}$ are zero modes of the quadratic action~\eqref{eq: color constant gauge field}, and we will exclude them in the semi-classical analysis. On the other hand, one can first see that $\xi_{jk,0;\pm}$  $(j>k)$ are also zero modes of the quadratic action~\eqref{eq: color constant gauge field}. To find the other zero modes among $\xi_{jk,n;\pm}$, we find the zeros of the $\lambda^{(j,k)}_{n,\pm}$ in addition to the trivial zero $n=0$. For $(j,k)$ such that $\nu_j\ne \nu_k$, the eigenvalue $\lambda_{n,\pm}^{(j,k)}$ has two non-trivial zeros denoted as $\omega^{(j,k)}_{\pm,\pm}$ in addition to the zero $n=0$,
\begin{align}
    \lambda_{n,+}^{(j,k)}\,=\,0\quad&:\quad  \omega^{(j,k)}_{+,\pm}\,=\, {1\over 2} \sqrt{\nu_j^2+\nu_k^2 \pm \sqrt{\big(\nu_j^2+\nu_k^2\big)^2-16\big(\nu_j^2-\nu_k^2\big)^2 }}\,, \label{eq: eigenvalue zero 1}\\
    \lambda_{n,-}^{(j,k)}\,=\,0\quad&:\quad \omega^{(j,k)}_{-,\pm}\,=\,- {1\over 2} \sqrt{\nu_j^2+\nu_k^2 \pm \sqrt{\big(\nu_j^2+\nu_k^2\big)^2-16\big(\nu_j^2-\nu_k^2\big)^2 }}\,. \label{eq: eigenvalue zero 2}
\end{align}
For $(j,k)$ with $\nu_j= \nu_k$, the eigenvalue $\lambda_{n,\pm }^{(j,k)}$ becomes simple
\begin{align}
    \lambda_{n,+}^{(j,k)}\,=\, n^2(n^2-\nu_j^2)\;\;,\qquad \lambda_{n,-}^{(j,k)}\,=\, n^2\,. 
\end{align}
Whence, $\lambda_{n,+ }^{(j,k)}$ has zero at $n=\nu_j$ and zeros with double root at $n=0$ while  $\lambda_{n,-}^{(j,k)}$ has one double root at $n=0$. Note that for the case of $\nu_j\ne \nu_k$ the zeros $\omega^{(j,k)}_{+,\pm}$ \eqref{eq: eigenvalue zero 1} and $\omega^{(j,k)}_{-,\pm}$ \eqref{eq: eigenvalue zero 2} are not integer in general, and $\xi_{jk,0;\pm}$ is the only zero mode among $\xi_{jk,n;\pm}$'s. This implies that the isometry $SU(N,N)/\mathbb{Z}_2$ of the AdS$_2$ background is broken in the non-trivial colored gravity backgrounds. For special values of $\nu$'s, the zeros $\omega^{(j,k)}_{\pm,\pm}$ \eqref{eq: eigenvalue zero 1}--\eqref{eq: eigenvalue zero 2} happen to become  integers so that the broken isometry can be enhanced.

\subsection{Lyapunov Exponents of the Colored Gravity}

The zeros of kernel in the quadratic action~\eqref{eq: color quadratic action diagonal}, or, equivalently, the poles in the propagator of the modes \eqref{eq: color propagator 1}, could appear as the Lyapunov exponent in the out-of-time-ordered correlator~(OTOC) of the matter coupled to the colored gravity~\cite{Sarosi:2017ykf,Jahnke:2019gxr,Narayan:2019ove,Yoon:2019cql}.\footnote{In this paper we assume that a matter field can be coupled to the colored gravity.} In the field theory with the Hamiltonian $H$, the Lyapunov exponent $\lambda_L$ can be defined by the exponential growth rate in real time $t$ of the (regularized) OTOC $F(t)$ of two operators $V(t)$ and $W(0)$~\cite{Shenker:2013pqa,Shenker:2014cwa}:\footnote{In chaotic system we expect the universal behaviour of the OTOC of typical operators.}
\begin{align}
	F_{\text{\tiny OTOC}}(t)\,\equiv\, \Tr\big[e^{-{\beta H\over 4}} \,V(t)\,e^{-{\beta H\over 4}}\,W(0)\,e^{-{\beta H\over 4}}\,V(t) \,e^{-{\beta H\over 4}}\,W(0)\big] \,\sim\, 1- \varepsilon e^{\lambda_L t} \,,  \label{eq: general otoc}
\end{align}
where $\varepsilon$ is proportional to ${1\over N^\alpha}$ in large-$N$ models for an appropriate $\alpha$ which is proportional to the Newton constant $G_N$ in the holographic dual gravity. The Lyapunov exponent $\lambda_L$ is bounded in the unitary QFTs \cite{Maldacena:2015waa}:
\begin{align}
    \lambda_L\,\leqq \, {2\pi \over \beta }\,. \label{eq: bound on chaos}
\end{align}
This bound is violated in the 3D higher spin gravities with finite number of higher spin fields~\cite{Perlmutter:2016pkf,David:2017eno,Yoon:2019cql,Datta:2021efl}. 
We will examine 
the bound violation issue in the 2D colored gravity.

The OTOC \eqref{eq: general otoc} can be evaluated by an appropriate analytic continuation of the Euclidean 4-point function $F_{\text{\tiny Eucl}}(u_1,u_2,u_3,u_4)$.\footnote{We will take the analytic continuation $u\;\to\; -{2\pi i \over \beta}t$ of $F_{\text{\tiny Eucl}}(u-\pi/2,u+\pi/2,0,\pi)$.} Assuming that the  4-point function is dominated by the OPE channel of which the intermediate operator is holographically dual to the graviton, one can approximate the 4-point function by the 2-point functions of the intermediate  graviton. In our case we assume that the OTOC of the boundary operators is dominated by the OPE channel of the (singlet and colored) gravitons and spin-1 mode. It can be approximated by their 2-point functions~\eqref{eq: color propagator 1} as 
\be
\label{eq: otoc contribution}
\begin{alignedat}{2}
&F_{\text{\tiny Eucl}}(u_1,u_2,u_3,u_4)\,=\,  \sum_{j=1}^N\sum_{n\in \mathbb{Z}/ \{ 0, \nu_j\}}\langle f_{jj,-n} f_{jj,n} \rangle \big[\delta_{f_{jj,-n}} \! G(1,2)\big]\big[\delta_{f_{jj,n}}G(3,4)\big] \\
&\hspace{10mm}+ \sum_{j=1}^N\sum_{n\in \mathbb{Z}/ \{ 0\}}\langle \phi_{jj,-n} \phi_{jj,n} \rangle \big[\delta_{\phi_{jj,-n}}G(1,2)\big]\big[\delta_{\phi_{jj,n}}G(3,4)\big]\\
&\hspace{10mm}+  \sum_{j>k}^N\ \sum_{n\in \mathbb{Z}/ \{0, \omega^{(j,k)}_{+,\pm} \}}\langle \xi_{jk,n;+}^\ast \xi_{jk,n;+} \rangle \big[\delta_{\xi_{jk,n;+}^\ast}G(1,2)\big]\big[\delta_{\xi_{jk,n;+} }G(3,4)\big]\\
&\hspace{10mm}+  \sum_{j>k}^N\ \sum_{n\in \mathbb{Z}/ \{0, \omega^{(j,k)}_{-,\pm} \}}\langle \xi_{jk,n;-}^\ast \xi_{jk,n;-} \rangle \big[\delta_{\xi_{jk,n;-}^\ast}G(1,2)\big]\big[\delta_{\xi_{jk,n;-} }G(3,4)\big]+\cdots\,, 
\end{alignedat}
\ee
where the soft mode eigenfunction, $\delta_{f_{jj,n}} G(1,2)\equiv {\delta G(u_1,u_2)\over \delta f_{jj,n}}$, corresponds to the OPE of the boundary matter operators with the soft mode $f_{jj,n}$, and it can be obtained by the infinitesimal transformation of the boundary-to-boundary 2-point function $G(u_1,u_2)$ of the boundary matter operator with respect to $f_{jj,n}$. Note that the ellipse denotes the contributions from conformal blocks of other intermediate operators, and we assume that they are subleading in the conformal block expansion. It is crucial to note that since the isometry of the background is modded out then  the mode corresponding to the isometry should be excluded in the summation in \eqref{eq: otoc contribution}. The summation over $n$ can be expressed as a contour integral around integer simple poles except for those isometry modes.\footnote{The form of the boundary-to-boundary propagator and its infinitesimal transformation is not known. Nevertheless, the result for the Lyapunov exponent is independent of their detailed form.} By changing the contour such a  contour integral is reduced to the residue at the zeros of the kernel in the quadratic action which was excluded in the original summation.\footnote{The pole at $n=0$ is subtle because the soft mode eigenfunction could vanish at $n=0$. However, the contribution from $n=0$ does not grow exponentially in time.} After the analytic continuation to real time, this gives the exponential growth of the OTOC:
\begin{align}
    F_{\text{\tiny OTOC}}(t)\,= \,& \sum_{j=1}^N e^{{2\pi  \nu_j \over \beta}t }+ \sum_{j>k} \bigg[e^{{2\pi  \omega_{+,+}^{(j,k)} \over \beta }t }+ e^{{2\pi  \omega_{+,-}^{(j,k)} t\over \beta}t }\bigg]+\cdots\,, 
\end{align}
where we included the terms which grow exponentially in real time $t$ up to prefactor.\footnote{For the detailed calculations of the OTOCs, see \cite{Narayan:2019ove}.} Therefore, the Lyapunov exponent of the colored gravity is given by  
\begin{align}
    \lambda_L \,=\, \max\bigg[\bigg\{ {2\pi \nu_N\over \beta} \bigg\}\,\cup\, \bigg\{ \text{Re}\bigg({2\pi \omega_{+,+}^{(j,k)}\over \beta}\bigg)\,\bigg|\, j>k\bigg\}\,\cup\, \bigg\{ \text{Re}\bigg({2\pi \omega_{+,-}^{(j,k)}\over \beta}\bigg)\,\bigg|\, j>k\bigg\}\bigg]\,, 
\end{align}
where we used $\nu_1\leqq \nu_2\leqq \ldots \leqq \nu_N$. When $\nu_N>1$, the Lyapunov exponent $\lambda_L$ is larger than ${2\pi \over \beta}$, and it violates the bound on chaos~\eqref{eq: bound on chaos}. The non-unitarity of the colored JT gravity, in particular, the instability of the spin-1 mode, is responsible for the violation of the bound on chaos. When $\nu_j=1$ for all $j=1,2,\ldots, N$ which correspond to the  AdS$_2$ background, we can see that the Lyapunov exponent $\lambda_L={2\pi \over \beta}$ saturates the bound although the instability of the spin-1 mode still persists even for the AdS$_2$ background. The exponential growth of the OTOC comes from (singlet and colored) gravitons, which are decoupled from the unstable spin-1 mode in the AdS$_2$ background. Therefore, in spite of the instability of the colored JT gravity, the bound on chaos~\eqref{eq: bound on chaos} seems to hold at the quadratic level of the boundary action. However, unlike the $sl(2,\mathbb{R})$ Schwarzian theory,  quantum corrections\footnote{See Ref.~\cite{Qi:2019gny} for the $sl(2,\mathbb{R})$ case.} to the Lyapunov exponent, which can be evaluated from the loop corrections to the propagators~\eqref{eq: color propagator 1} of gravitons and spin-1 mode induced by  interaction terms in the boundary effective action~\eqref{f c Sch}, might violate the bound on chaos even for the AdS$_2$ background because of the interaction among the unstable spin-1 mode and the gravitons.

%%%%%%%%%%%%%%%%%%%%%%%%%%%%%%%%%%%%%%%%%%%%%%%%%%%%%%%%%%%%%%%%%%%%%%
%%%%%%%%%%%%%%%%%%%%%%%%%%%%%%%%%%%%%%%%%%%%%%%%%%%%%%%%%%%%%%%%%%%%%%
\section{Rainbow AdS$_2$}
\label{sec:rainbow}
%%%%%%%%%%%%%%%%%%%%%%%%%%%%%%%%%%%%%%%%%%%%%%%%%%%%%%%%%%%%%%%%%%%%%%
%%%%%%%%%%%%%%%%%%%%%%%%%%%%%%%%%%%%%%%%%%%%%%%%%%%%%%%%%%%%%%%%%%%%%%

So far, we have considered the asymptotic AdS$_2$ condition~\eqref{asym_cond} together with the gauge condition~\eqref{eq: color FG gauge},
\be
	A(r,u)\,=\,b^{-1}(r)\big(a(u)\rd u + \rd \big)b(r)\,=\, \begin{pmatrix}
	i\bmcJ(u) & -\bmcL(u) e^{-r} \\
	\bmI e^r & i\bmcJ(u)\\
	\end{pmatrix},\label{eq: the form of A ru}
\ee
where quantities $\bmcJ(u)$ and $\bmcL(u)$ are defined by \eqref{JL}.\footnote{In the 3D higher-spin Chern-Simons  gravity, $\bmcL$ is analogous to ``charge'' while ``source'' is located at the position of $\bmI$ in Eq.~\eqref{eq: the form of A ru}~\cite{Gutperle:2011kf,Ammon:2011nk,Castro:2011fm,Henneaux:2013dra}.} It allows the background only for the spin-$2$ singlet at asymptotic infinity $r\to \infty$. As a result, the trace part of $\bmcL$ containing both singlet and colored spin-2 modes appears in the boundary effective action~\eqref{eq: boundary action zero temp}. 

We may also consider an ansatz for the connection where the colored spin-$2$ modes also acquire backgrounds at asymptotic infinity,
\be
\begin{alignedat}{2}
	a(u)\,=\,&  \big(L_1 +\mathcal{L}(u)\, L_{-1}
	+\mathcal{M}(u)\,L_0\big)\otimes \bm{I}
	\\
	 +& \big(X_A\,L_{1}+\mathcal{K}_A(u)\,L_{-1}+\mathcal{N}_A(u)\,L_0\big)\otimes\bm \bT^A +i\,\cJ_A(u)\, I \otimes \bT^A\,,
\end{alignedat}
\ee
where the backgrounds of the colored spin-2 modes
are given by the constants $X_A$,
and we have also introduced $\mathcal{M}(u)$
and $\mathcal{N}_A(u)$ 
to make the ansatz sufficiently general.
The above ansatz can be represented as a matrix form,
\begin{equation}
    a(u)=\begin{pmatrix}
    \,i\,\bm{\cW} & - \bm{\cL} \\
    \bm Z & \,i\,\bm{\cW}^\dagger 
    \end{pmatrix}, 
    \label{eq: general form of gauge field a}
\end{equation}
where $\bm \cL(u)$ and $\bm \cW(u)$ 
are matrix functions
while $\bm Z$ is a constant matrix:
\be
\begin{alignedat}{2}
\bm{\mathcal L}(u)\,=\,& \mathcal L(u)\,\bm I + \mathcal K_A(u)\,\bT^A\,,
\\
\bm\cW(u)\,=\,& 
\cJ_A(u)\,\bT^A
-\frac{i}2\left(\cM(u) \,\bmI+ \cN_A(u)\, \bT^A\right),
\\
\bm Z\,=\, & \bm I+X_A\,\bT_A\;.
\end{alignedat}
\ee
The choice of constant $\bm Z$ (i.e. $X_A$) corresponds to a stationary background at asymptotic infinity. Since this mimics the ``rainbow solutions'' of the 3D colored gravity \cite{Gwak:2015vfb} we will also keep this term in the 2D colored gravity. However, the main difference in the present case is that the 2D colored gravity action has no  potential term for the variable $X_A$ to be minimized. Furthermore, unlike the ansatz \eqref{eq: the form of A ru} with $\bm Z=\bmI$  in the present case we need to check whether one can impose  the gauge  $\bm \cW=\bm \cW^\dagger$ (i.e. $\mathcal{M}=\mathcal{N}_A=0$) by using the residual gauge symmetry.

\subsection{Rainbow Effective Action}

Let us first consider a  gauge transformation of the rainbow solution \eqref{eq: general form of gauge field a}:
\be
\label{const_gauge_tr}
    a'\,=\,U^{-1}\,a\,U\,=\,
    \begin{pmatrix}
    i\,\bm \cW' & -\bm \cL' \\
    \bm Z' & i\,\bm \cW'^\dagger\\
    \end{pmatrix},
    \qquad
    U\,=\, \begin{pmatrix}
    \bm A & \bm B\\
    \bm C & \bm D\\
    \end{pmatrix},
\ee
where $U$ is  taken to be a  constant $SU(N,N)$ matrix~\eqref{SU(N,N)}. The resulting  transformed components $\bm\cW'$, $\bm\cL'$,  $\bm Z'$   are collected in Appendix \bref{app:rainbow},  see \eqref{this_eq1}--\eqref{this_eq4}. 
In general, such a transformation modifies the asymptotic behaviour of the connection $A(r,u)$ \eqref{eq: color FG gauge}. To keep $\bm Z'$ constant that guarantees a stationary background at asymptotic infinity one chooses a constant gauge parameter $U$ with $\bm B= \bm C=0$ and $\bm A\bm D=\bmI$ that yields  the following  transformation of $\bm Z$ inherited from the relation \eqref{this_eq4}:
\be
\bm Z' \,=\, \bm A^\dag \bm Z \bm A\,.
\ee
Then, one may consider all $\bm A\in GL(N,\mathbb C)$ such that there  are left $N+1$ different $GL(N,\mathbb C)$ orbits in the space of $\bm Z$ with the representatives 
\begin{align}
    \bmZ_{(p,q)}\,=\, \text{diag}(\;\overbrace{1,1,\cdots, 1}^p\;,\; \overbrace{-1,-1,\cdots, -1}^q\;) \hspace{8mm} p+q=N\,. \label{eq: representation z}
\end{align}
Taking into account the $\mathbb Z_2$ quotient which modes out the overall sign, there will be $[(N+1)/2]$ different orbits left. This reproduces the equation for $\bm Z_{(p,q)}$ found in the context of the AdS$_3$ colored gravity \cite{Gwak:2015vfb}, 
\be
\bm Z_{(p,q)}^2 \,=\, \bmI\,.
\ee
Choosing $\bm Z_{(p,q)}$ \eqref{eq: representation z} we introduce the asymptotic rainbow-AdS$_2$ condition
\begin{equation}
	\left.(A-A_{\text{\tiny rAdS}})\right|_{\partial \mathcal{M}_2}\,=\,\cO(1)\,,\label{eq: asymptotic rainbow ads condition}
\end{equation}
where the rainbow-AdS$_2$ background is defined by the connection 
\be
\label{rain_back}
A_{\text{\tiny rAdS}}= L_1 \otimes \bm Z_{(p,q)}+ L_{-1} \otimes \bmcL_0\;,
\qquad
\bmcL_0 = \mbox{const.} 
\ee
In general, the $SU(N)$ color gauge symmetry is broken by the  asymptotic rainbow-AdS$_2$  condition. To see this, let us  consider a color gauge transformation $a'=h^{-1} a h + h^{-1} \dot{h}$ with the gauge parameter
\be
    U\,=\, \begin{pmatrix}
    \bm V(u) & \bm 0 \\
    \bm 0 & \bm V(u)\\
    \end{pmatrix},\qquad\text{where}\quad \bm V(u) \in SU(N)\,,
\ee
under which $\bm Z_{(p,q)}$  transforms  as
\be
    \bm Z'_{(p,q)}\,=\, \bm V^{-1} \bm Z_{(p,q)} \bm V \,. 
\ee
Hence, unless $\bm Z_{(N,0)}=\bmI$, the $SU(N)$ color gauge symmetry is spontaneously broken to the subgroup  $U(p)\otimes U(q)/U(1)$. 

This also affects the gauge condition that can be chosen by using residual gauge symmetry keeping the asymptotic rainbow-AdS$_2$ condition~\eqref{eq: asymptotic rainbow ads condition}. To this end, let us consider the residual gauge transformation of $a(u)$ \eqref{eq: general form of gauge field a} by the gauge parameter $U(u)$ of the form 
\begin{align}
    U(u)\,=\, \begin{pmatrix}
    \bmI & \bm h(u)\\
    \bm 0 & \bmI \\
    \end{pmatrix}
    \quad \text{for some} \quad \bm h^\dagger = \bm h \,.
\end{align}
Then, we find that  $\bm \cW$ transforms as 
\begin{equation}
    \bm \cW' = \bm\cW +i\,\bmh \, \bmZ \, \equiv\,\bm\cW+ i\,\begin{pmatrix}
    \bm h_{(p,p)} & -\bm h_{(p,q)} \\
    \bm h_{(q,p)} &  - \bm h_{(q,q)}\\
    \end{pmatrix}\,,
\end{equation}
where we used the block-matrix notation \eqref{M_block} for the matrix $\bm h$:
the Hermiticity of $\bm h$ implies 
$\bm h_{(p,p)}^\dagger=\bm h_{(p,p)}$ 
and $\bm h_{(q,q)}^\dagger=\bm h_{(q,q)}$ 
while $\bm h_{(p,q)}^\dagger=\bm h_{(q,p)}$\,.
Notice that unlike the case $\bm Z=\bmI$ one cannot choose a gauge where $\bm\cW=\bm \cW^\dagger$
because $\bm\cW_{(p,q)}^\dagger-\bm\cW_{(q,p)}$
is invariant under the transformation.
Instead, we can set $\bm\cW_{(p,q)}^\dagger+\bm\cW_{(q,p)}$
to zero, and obtain
\be
\label{eq: colored gauge field a}
    \bm \cW \,=\, \begin{pmatrix}
    \bm \cJ_{(p,p)} & -i\,\bm \cJ_{(p,q)} \\
    -i\,\bm \cJ_{(q,p)} & \bm \cJ_{(q,q)} \\
    \end{pmatrix}\, ,
\ee
with $\bm\cJ_{(p,p)}^\dagger=\bm\cJ_{(p,p)}$,
$\bm\cJ_{(q,q)}^\dagger=\bm\cJ_{(q,q)}$ 
and $\bm\cJ_{(p,q)}^\dagger=\bm\cJ_{(q,p)}$\,.

Using the parameterization \eqref{g decomp} for the group element $g$ in $a=g^{-1}\,\dot g$ and taking $a$ in the asymptotic rainbow-AdS$_2$ condition~\eqref{eq: asymptotic rainbow ads condition} in the form \eqref{eq: colored gauge field a}  we find
\be
\label{eq: rainbow fdot}
\begin{alignedat}{2}
\dot{\bm f}\,=\,&\bm b\,\bm d\,\bm Z\,\bm d^{-1}\,\bm b\,,  \\
\bm \cL\,=\,&\dot{\bm e}-\bm e\,\bm Z\,\bm e+ 
i\left(\bm \cW\, \bme - \bm e\, \bm \cW^\dagger\right),
\\
i\,\bm \cW\,=\,& \bm e\,\bm Z+\bm d^{-1}\,\dot{\bm d} -\bm d^{-1}\,\dot{\bm b}\,\bm b^{-1}\,\bm d \,.
\end{alignedat}
\ee
Then, the boundary effective action~\eqref{eq: boundary action zero temp} is  given again by
\be
\label{eq: bdy action rainbow}
\begin{alignedat}{2}
    S_{tot}\,=\,& {\kappa \gamma \over 2 N} \int \rd u\; \tr (a^2)\,=\,-{\kappa \gamma \over N} \int \rd u\; \tr \bigg[\bmcL\,\bmZ + \half
    \left(\bm \cW^2 +\bm \cW^\dagger{}^2\right)\bigg]\\
    \,=\,&-{\kappa \gamma \over  N} \int \rd u\;\tr \bigg[\bmb^{-1} \ddot{\bmb} -2(\bmb^{-1}\dot{\bmb})^2 - (\bmd^{-1}\dot{\bmd})^2 - \dot{\bmd}\bmd^{-1}(\bmb^{-1}\dot{\bmb}-\dot{\bmb}\,\bmb^{-1}) \bigg]\ .
\end{alignedat}
\ee
Remarkably, the last  expression is formally identical to that one in the case  $\bm Z=\bmI$ \eqref{mat Sch}.
However, when $\bm Z\neq  \bmI$, there exists a crucial difference in the physical property compared to the $\bm Z=\bmI$ case. 

In order to see this point, we expand the action \eqref{eq: bdy action rainbow} perturbatively around the branch $\bmb^2\,=\, \dot{f} \bmI$,  i.e. 
\begin{align}
    \bm b\,=\, \dot{f}^{1\over 2} \big(\bmI + \bm\eta\big)\,,
\end{align}
where $\bm\eta$ stands for a fluctuation
and $f\equiv \tr (\bmZ \bmf) $
is the singlet graviton mode
in the symmetry-broken phase. In this way one obtains the perturbative expansion of the action not only in $\bm\eta$, but also in $\bm k\equiv \bm f-f\,\bm Z$ (or, equivalently, in $\bm \alpha \equiv \dot f^{-1}\dot{\bm k}$) and $\bm \phi=-i\,\log \bm d$. The first-order solution is found to be
\begin{equation} 
\bm\eta_{\rm d}
+i\,\bm\phi_{\rm o}
=\bm \alpha+\cO((\bm\eta,\bm\alpha,\bm\phi)^2)\,,
\end{equation}
where 
\be
\ba{l}
\bm\eta_{\rm d}
=\half(\bm \eta+\bm Z\,\bm\eta\,\bm Z)\,,
\vspace{2mm}
\\
\bm\phi_{\rm o}=\half(\bm \phi-\bm Z\,\bm\phi\,\bm Z)\,,
\ea
\ee
are diagonal and off-diagonal block parts of $\bm\eta$ and $\bm\phi$, respectively. 
The remaining block parts 
\be
\ba{l}
\bm\eta_{\rm o} =\half(\bm \eta-\bm Z\,\bm\eta\,\bm Z)\,,
\vspace{2mm}
\\
\bm\phi_{\rm d}=\half(\bm \phi+\bm Z\,\bm\phi\,\bm Z)\,
\ea
\ee
are left independent. Using the following perturbative relations
 \be
 \label{eq: rainbow-AdS perturbative cal}
 \begin{alignedat}{2}
     {1\over 2}\big(\bm b^{-1}\,\dot{\bm b}
     -\dot{\bm b}\, \bm b^{-1}\big)
     \,=\,&  \mathcal{O}\big((\bm\eta,\bm\alpha,\bm\phi)^2\big)\ ,
     \\
      {1\over 2}\big(\bm b^{-1}\,\dot{\bm b}
     +\dot{\bm b}\, \bm b^{-1}\big)
     \,=\,&{\ddot f\over 2 \dot{f} }\,\bm I
     + 
    %  \begin{pmatrix}
    %      {1\over 2} \bm\dot{\alpha}_{(p,p)} &  \bm\dot{\eta}_{(p,q)}\\
    %      \bm\dot{\eta}_{(q,p)} & -{1\over 2}\bm\dot{\alpha}_{(q,q)}\\
    %  \end{pmatrix}
     \frac{\{\bm Z,\dot{\bm \alpha}\}}4
     +\frac{\dot{\bm\eta}_{\rm o}}2
     + \mathcal{O}\big((\bm\eta,\bm\alpha,\bm\phi)^2\big) \ ,
\end{alignedat}
\ee
we find 
\be
 \label{eq: rainbow-AdS perturbative cal2}
\begin{alignedat}{2}
    &-\tr\left[\bm \cL\,\bm Z+\frac12\left(\bm \cW^2+\bm\cW^\dagger{}^2\right)\right]\\
    \,=\,& {N\over 4} \bigg({\ddot{f}\over \dot{f}}\bigg)^2 +{1\over 4}\, \tr \big(\bm Z\,\dot{\bm \alpha}\big)^2-\tr \big(
    \dot{\bm \phi}_{\rm d}^2\big)
    +\tr \big( 
    \dot{\bm\eta}_{\rm o}^2 \big) +\cO\big((\bm\eta,\bm\alpha,\bm\phi)^3\big)+
    (\textrm{total derivative})\ ,
\end{alignedat}
\ee
where we used $\tr(\bm k\,\bm Z)=0$\,. 

Two remarks are in order.
First, 
in the off-diagonal block modes,
the derivatives of 
the spin-two modes $\bm k$
are expressed by $\bm\phi_{\rm o}$ rather than $\bm\eta_{\rm o}$.
At the same time, $\bm\eta_{\rm o}$
plays the role of spin-1 modes.
This means that the spin-connection of 
the spin-two modes
and the spin-one modes
are interchanged in the symmetry broken part.
This is in agreement
with the Higgs-like mechanism
of the colored JT gravity in the bulk \cite{Alkalaev:2022szj}.
Second, compared to the $\bmZ=\bmI$ case \eqref{k exp}, the overall signs of the off-diagonal block modes, $\bm\alpha_{\rm o}$ and $\bm \eta_{\rm o}$, are flipped (up to total derivatives).
Therefore, the off-diagonal block spin-2 modes $\bm\alpha_{\rm o}$ gets to be unstable,
while the off-diagonal block spin-1 modes $\bm \eta_{\rm o}$ becomes stable because of the broken color symmetry. We will also reach the same conclusion in Section~\bref{sec: perturbative expansion rainbow} by the perturbative analysis at finite temperature.

\subsection{Asymptotic Rainbow Symmetry}

Let us now consider the asymptotic symmetry for the rainbow AdS$_2$. As in Section~\bref{sec: asymptotic symmetry}, we consider the infinitesimal gauge transformation of the connection $a(u)$ \eqref{eq: colored gauge field a}:
\be  
\label{rain_gauge_tran} 
\delta a(u)\,=\, \dot{h} + [a(u),h]\;,
    \qquad
    h(u)\,=\,\begin{pmatrix}
    \bml & \bmn \\
    \bmm & -\bml^\dag \\
    \end{pmatrix}, \quad \text{where}\;\; \bml\,\equiv\,\bms+i \bmt\,,
\ee
see \eqref{eq: transf of a rainbow}. Imposing the asymptotic rainbow-AdS$_2$ condition $\delta a(u)=0$ we find the components of the connection and the parameter are related by \eqref{eq: asym rads sol 1}--\eqref{eq: asym rads sol 5} given in terms of the block-diagonal decomposition \eqref{M_block}. 
Unlike the case $\bm Z=\bmI$, one can solve the asymptotic rainbow-AdS$_2$ condition for
the Hermitian matrices $\bmn$,
 $\bms_{(p,p)}$, $\bms_{(q,q)}$ and 
the complex matrix  $\bmt_{(p,q)}$. Hence, the asymptotic rainbow-AdS$_2$ symmetry can be parametrized by the Hermitian matrices $\bmm$,
$\bmt_{(p,p)}$, $\bmt_{(q,q)}$
and 
the complex $\bms_{(p,q)}$.  We conclude that  the rainbow-AdS$_2$ has $2N^2-1$ boundary degrees of freedom
(taking into account the traceless condition $\tr(\bm t_{(p,p)}+\bm t_{(q,q)})=0$), which is the same as in the case $\bmZ=\bmI$. However, note  a different organization of the degrees of freedom: in the rainbow-AdS$_2$ case, there  are off-diagonal $\bms_{(p,q)}$  boundary degrees of freedom instead of  $\bmt_{(p,q)}$   in the $\bmZ=\bmI$ case.\footnote{Note that for $\bmZ=\bmI$ the $(p,q)$-decomposition   \eqref{M_block} is superfluous and can be done only to compare with the rainbow-AdS$_2$ case.}  

The residual transformation of the connection  $a(u)$ \eqref{eq: transf of a rainbow} defines  the asymptotic rainbow-AdS$_2$ transformation in the block-diagonal notation:
\begin{align}
&\delta \bmcJ_{(p,p)}\,=\, \cD_{\cJ_{(p,p)}} \bmt_{(p,p)} + {i\over 2}\,\big([\bmcL,\bmm]\big)_{(p,p)} -i\big(\bmcJ_{(p,q)} \bms_{(q,p)}- \bms_{(p,q)} \bmcJ_{(q,p)}\big)\,,\label{eq: rainbow asymptotic transf 1}\\
&\delta \bm\cJ_{(q,q)}\,=\, \cD_{\cJ_{(q,q)}} \bmt_{(q,q)} + {i\over 2}\,\big([\bmcL,\bmm]\big)_{(q,q)} -i\big(\bm\cJ_{(q,p)} \bms_{(p,q)}- \bms_{(q,p)} \bmcJ_{(p,q)}\big)\,,\label{eq: rainbow asymptotic transf 2}\\
&\begin{alignedat}{2}
\delta \bm\cJ_{(p,q)}\,=\,&\left(\dot{\bms}-\half\bmcL\, \bmm-\half\bmm\, \bmcL\right)_{(p,q)} + i\big(\bm\cJ_{(p,p)}\, \bms_{(p,q)}-\bms_{(p,q)}\, \bm\cJ_{(q,q)})
\vspace{2mm}
\\
&+ i\big(\bm\cJ_{(p,q)}\, \bmt_{(q,q)}-\bmt_{(p,p)}\, \bm\cJ_{(p,q)}) \,,\label{eq: rainbow asymptotic transf 3}
\end{alignedat}\\
&\begin{alignedat}{2}
\delta \bmcL\,=\,& - \dot{\bmn} -\{\bmcL, \bms\}+i\, [\bmcL, \bmt] 
\vspace{2mm}
\\
&
 -  i\begin{pmatrix}
    \bm\cJ_{(p,p)}\, \bmn_{(p,p)}-\bmn_{(p,p)}\, \bm\cJ_{(p,p)} &\;\;\;\;  \bm\cJ_{(p,p)}\, \bmn_{(p,q)}-\bmn_{(p,q)}\, \bm\cJ_{(q,q)}\\
    \bm\cJ_{(q,q)}\, \bmn_{(q,p)}-\bmn_{(q,p)}\, \bm\cJ_{(p,p)} &\;\;\;\; \bm\cJ_{(q,q)}\, \bmn_{(q,q)}-\bmn_{(q,q)}\, \bm\cJ_{(q,q)}\\
    \end{pmatrix}
\vspace{2mm}
\\
&- \begin{pmatrix}
    \bm\cJ_{(p,q)}\, \bmn_{(q,p)}+\bmn_{(p,q)}\, \bm\cJ_{(q,p)} &\;\;\;\; \bm\cJ_{(p,q)}\, \bmn_{(q,q)}+\bmn_{(p,p)}\, \bm\cJ_{(p,q)}\\
    \bm\cJ_{(q,p)}\, \bmn_{(p,p)}+\bmn_{(q,q)}\, \bm\cJ_{(q,p)} &\;\;\;\; \bm\cJ_{(q,p)}\, \bmn_{(p,q)}+\bmn_{(q,p)}\, \bm\cJ_{(p,q)}\\
    \end{pmatrix}\ .\label{eq: rainbow asymptotic transf 5}
\end{alignedat}
\end{align}
Here, as discussed before,  parameters    $\bmn$, $\bms_{(p,p)}$, $\bms_{(q,q)}$, $\bmt_{(p,q)}$ should be expressed in terms of independent parameters $\bmm$, $\bmt_{(p,p)}$, $\bmt_{(q,q)}$, $\bms_{(p,q)}$  via \eqref{eq: asym rads sol 1}--\eqref{eq: asym rads sol 5}. This is the rainbow-extension of the asymptotic AdS$_2$ symmetry in Section~\bref{sec: asymptotic symmetry} and in Ref.~\cite{Joung:2017hsi}. In the rainbow-AdS$_2$ case, the matrix $\bm \cJ_{(p,q)}$ is the generator of the broken symmetry. 

Furthermore, compared to the asymptotic AdS$_2$ transformations \eqref{eq: color asymptotic ads transf 1} of $\bm\cJ $ and $\bmcL$ in the $\bm Z=\bmI$ case, there are simple at first sight  extra contributions: the last two terms in \eqref{eq: rainbow asymptotic transf 1}--\eqref{eq: rainbow asymptotic transf 2} and the last term in \eqref{eq: rainbow asymptotic transf 5}. However, due to the asymptotic rainbow-AdS$_2$ constraints ~\eqref{eq: asym rads sol 1}--\eqref{eq: asym rads sol 5}, the resulting asymptotic transformation of $\bmcJ$ and $\bmcL$ parameterized by $\bmm$,  $\bmt_{(p,p)}$, $\bmt_{(q,q)}$, $\bms_{(p,q)}$ turn out to be much more complicated.

The asymptotic rainbow-AdS$_2$ transformations \eqref{eq: rainbow asymptotic transf 1}--\eqref{eq: rainbow asymptotic transf 5} are not enough to understand the consequence of the broken color symmetry and the organization of the resulting boundary degrees of freedom. Therefore, we will evaluate the boundary effective action for the rainbow-AdS$_2$ case up to quadratic order as in Section~\ref{sec: quadratic action}. To this end, we first consider the smooth rainbow-AdS$_2$ background $a_0$ \eqref{rain_back} given by
\begin{align}
    a_0\,=\, \begin{pmatrix}
    \bm 0 & - \bmcL_0\\
    \bmZ & \bm 0\\
    \end{pmatrix}\,, 
\end{align}
where $\bmcL_0$ is a constant matrix. By means of the $u$-independent gauge transformation
\begin{align}
    \begin{pmatrix}
    \bm U^{-1} \bmZ & \bm 0\\
    \bm 0 & \bmZ \bm U^{-1}\\
    \end{pmatrix}a_0\begin{pmatrix}
    \bmZ\bm U  & \bm 0\\
    \bm 0 & \bm U \bmZ\\
    \end{pmatrix}\,=\,& \begin{pmatrix}
    \bm 0 & - \bm U^{-1}\bmZ\bmcL_0 \bm U \bmZ\\
    \bmZ & \bm 0
    \end{pmatrix},
\end{align}
with constant  $\bm U\,\in\, U(N)$  one can also diagonalize\footnote{We demand that the matrix $a_0^2$ instead of $a_0$ is Hermitian. It follows that  the boundary effective action in \eqref{eq: bdy action rainbow} given by $\tr a^2$ is real. Therefore, the matrix $\bmZ\bmcL_0$ is Hermitian.} the constant matrix $\bmZ \bmcL_0$:
\begin{align}
    \bmZ \bmcL_0\,=\, \text{diag}\big( \lambda_1\;,\;\cdots\;,\;  \lambda_N \big)\,. 
\end{align}
From the trivial holonomy condition discussed in Section~\ref{sec: smooth background}, the eigenvalues of $\bmZ \bmcL_0$ are determined to be
\begin{align}
    \lambda_j\,=\, {\pi^2\nu_j^2\over \beta^2}\;,\hspace{8mm}\mbox{where}\quad \nu_j\in \mathbb{Z}_+\,\quad\text{and}\quad \nu_1\leqq \cdots\leqq \nu_N\ .
\end{align}
Here all $\nu_j$'s are either even for $\text{Hol}(A)\,=\,I_{2N}$ or odd for $\text{Hol}(A)\,=\,-I_{2N}$. Note that the energy of the constant solution $a_0$ in \eqref{eq: energy of constant background} 
\begin{align}
\label{energy_rain}
    E_0\,=\, {\beta \kappa \gamma\over 2 N} \, \tr a_0^2\,=\, - {\beta\kappa\gamma\over N} \;\tr \big(\bmZ \bmcL_0\big) \,=\,-{\pi^2\kappa \gamma\over \beta N }\sum_{j=1}^N \nu_j^2\ . 
\end{align}
Note that the value of the energy
is not affected by
the color symmetry breaking,
which one could expect from
the fact that the colored JT gravity
does not have a potential term for the colored spin-2 modes as opposed
to its 3D analog \cite{Alkalaev:2022szj}.
In what follows,  we will focus on the highest-energy  constant solution given by
\begin{align}
    a_0\,=\, \begin{pmatrix}
    \bm 0 & - {\pi^2\over \beta^2} \bmZ\\
    \bmZ & \bm 0\\
    \end{pmatrix},  \label{eq: rainbow ads background}
\end{align}
for the sake of simplicity.

%%%%%%%%%%%%%%%%%%%%%%%%%%%%%%%%%%%%%%%%%%%%%%%%%%%%%%%%%%%%%%%%%%%
\subsection{Perturbative Expansion }
\label{sec: perturbative expansion rainbow}
%%%%%%%%%%%%%%%%%%%%%%%%%%%%%%%%%%%%%%%%%%%%%%%%%%%%%%%%%%%%%%%%%%%

As in Section~\ref{sec: quadratic action} we consider a small fluctuation around the  rainbow-AdS$_2$ background \eqref{eq: rainbow ads background} and evaluate the boundary effective  action in the quadratic approximation. Similarly, we  expand  the gauge parameter $U(u)$ \eqref{gauge_p_exp} with respect to  small fluctuations $h(u)$ and $k(u)$ given by \eqref{gauge_p_hk}. A fluctuation around the  background~\eqref{eq: rainbow ads background} can be parameterized by the gauge transformation of $a_0$ by $U(u)$ \eqref{eq: expansion of a color}. At order $\mathcal{O}(\epsilon)$, the asymptotic AdS$_2$ condition and the gauge condition~~\eqref{eq: colored gauge field a} allow to express block-diagonal components of $\bms$, $\bmt$, $\bmn$ in terms of the block-diagonal components of $\bmm$ as 
\be
\label{eq: rainbow ads 1st order sol1}
\bms_{(p,p)}\,=\,-\frac12\dot{\bmm}_{(p,p)}\,,
\quad
\bms_{(q,q)}\,=\,\frac12 \dot{\bmm}_{(q,q)}\,,
\quad
\bmt_{(p,q)}\,=\,{i\over 2} \dot{\bmm}_{(p,q)}\,,
\quad
\bmt_{(q,p)}\,=\,-{i\over 2} \dot{\bmm}_{(q,p)}\,,
\ee
\be
\bmn_{(p,p)}\,=\,
	-{1\over 2} \ddot{\bmm}_{(p,p)}-{\pi^2\over \beta^2}\bmm_{(p,p)}\;,
	\quad \bmn_{(p,q)} \,=\, {1\over 2} \ddot{\bmm}_{(p,q)}+{\pi^2\over \beta^2}\bmm_{(p,q)}
\;,
\ee
\be
\label{eq: rainbow ads 1st order sol4}
\bmn_{(q,p)} \,=\, 
{1\over 2} \ddot{\bmm}_{(q,p)}+{\pi^2\over \beta^2}\bmm_{(q,p)}\;,
\quad
	\bmn_{(q,q)}\,=\, -{1\over 2} \ddot{\bmm}_{(q,q)}-{\pi^2\over \beta^2}\bmm_{(q,q)} \;. 
\ee
These relations are obtained from  \eqref{eq: asym rads sol 1}--\eqref{eq: asym rads sol 5} by choosing $\bmcL_0={\pi^2\over \beta^2}\bmZ$ and $\bm\cW_0=0$. By their means one can perturbatively express $\bmcL$, $\bm \cW$  \eqref{eq: colored gauge field a} in terms of $\bmm$, $\bmt_{(p,p)}$, $\bmt_{(q,q)}$, $\bms_{(p,q)}$. Expanding $\bmcL$  and $\bm\cW$ with respect to $\epsilon$,
\be
\bmcL=\bmcL_0 + \epsilon\, \bmcL^{(1)}+\epsilon^2\, \bmcL^{(2)}+...,
\qquad
\bm\cW=\epsilon\, \bm\cW^{(1)}+\epsilon^2\, \bm\cW^{(2)}+...,
\ee
we can similarly expand  the effective action \eqref{eq: bdy action rainbow} up to a quadratic order,
\be
\ba{l}
\label{eq: bdy action rainbow2}
\dps
S_{tot} = {\kappa \gamma \over 2 N} \int \rd u\; \tr (a^2)\,=\,-{\kappa \gamma \over N} \int \rd u\; \tr \bigg[\bmcL \,\bmZ + \half
\left(\bm \cW^2 +\bm\cW^\dagger{}^2\right)\bigg]
\vspace{3mm}
\\
\dps
\hspace{8mm}= -{\kappa \gamma \over  N} \int \rd u\;\tr(\bmcL_0\,\bmZ)-{\epsilon\kappa \gamma \over  N} \int \rd u\;\tr  (\bmcL^{(1)}\,\bmZ)
\vspace{3mm}
\\
\dps
\hspace{20mm}
-{\epsilon^2\kappa \gamma \over  N} \int \rd u\;\tr 
\bigg[\bmcL^{(2)}\bmZ+ \half\left(\bm \cW^{(1)}{}^2 +\bm \cW^{(1)}{}^\dagger{}^2\right) \bigg] + \cO(\epsilon^3)\,.
\ea
\ee
The leading term here is the energy of the background $E_0=- {\pi^2\kappa \gamma \over \beta}$, see \eqref{energy_rain}, the  first-order term vanishes similarly to \eqref{trL1}, the second-order terms can be explicitly calculated according to \eqref{L1}--\eqref{L2}. Finally, using \eqref{eq: rainbow ads 1st order sol1}--\eqref{eq: rainbow ads 1st order sol4} one obtains the boundary effective action \eqref{eq: bdy action rainbow2} in terms of $\bmm$, $\bmt_{(p,p)}$, $\bmt_{(q,q)}$, $\bms_{(p,q)}$:
\be
\label{last_act}
\begin{alignedat}{2}
S_{tot}\,=\,& E_0 + {\kappa \gamma \epsilon^2 \over N} \int \rd u\; \tr \bigg[ {1\over 4} \, \ddot{\bmm}_{(p,p)}^2 - {\pi^2\over \beta^2} \dot{\bmm}_{(p,p)}^2+ {1\over 4} \, \ddot{\bmm}_{(q,q)}^2 - {\pi^2\over \beta^2} \dot{\bmm}_{(q,q)}^2\bigg]\\
& - {2\kappa \gamma \epsilon^2 \over N} \int \rd u\; \tr \bigg[ {1\over 4} \, \ddot{\bmm}_{(p,q)}\ddot{\bmm}_{(q,p)} - {\pi^2\over \beta^2} \dot{\bmm}_{(p,q)}\dot{\bmm}_{(q,p)} \bigg]\\
&- {\kappa \gamma \epsilon^2 \over N} \int \rd u\; \tr \bigg[   \dot{\bmt}_{(p,p)}^2+\dot{\bmt}_{(q,q)}^2  \bigg] + {2\kappa \gamma \epsilon^2 \over N} \int \rd u\; \tr \bigg[   \dot{\bms}_{(p,q)}\dot{\bms}_{(q,p)}  \bigg]+ \mathcal{O}\big(\epsilon^3\big)\ .
\end{alignedat}
\ee

One can evaluate the mode expansion of the quadratic boundary action along the lines of Section \bref{sec:modes}, see \eqref{eq: color quadratic action diagonal}. First, it is easy to see that the $(p^2+q^2-1)$ modes $\bmt_{(p,p)}$ and $\bmt_{(q,q)}$ have the same quadratic action as that of the unstable spin-1 mode in the colored JT graviton~\eqref{eq: color quadratic action diagonal}. However, the $2pq$ modes, $\bms_{(p,q)}=\big(\bms_{(q,p)}\big)^\dag$, also have the same quadratic action of $\bmt_{(p,p)}$ and $\bmt_{(q,q)}$ but with the opposite overall sign. Therefore, $\bms_{(p,q)}$ corresponds to the stable spin-1 mode. Furthermore, the $p^2+q^2$ modes $\bmm_{(p,p)}$ and $\bmm_{(q,q)}$ have the same quadratic action as that of the stable colored graviton~\eqref{eq: color quadratic action diagonal} while $2pq$ modes, $\bmm_{(p,q)}=\big(\bmm_{(q,p)}\big)^\dag$, have the same one but with the opposite overall sign. Due to this opposite sign, $\bmm_{(p,q)}$ is turned out to be the unstable spin-2 mode. The broken color symmetry made the broken part of the spin-2 unstable while the broken part of the spin-1 becomes stable. These results are summarized in Table~\ref{tab:mode}, which are consistent with the perturbative analysis in~\eqref{eq: rainbow-AdS perturbative cal2} 
and the Higgs-like
mechanism
of the colored JT gravity
in the bulk
\cite{Alkalaev:2022szj}.
A similar phenomenon has been observed in the Higgs-like mechanism of the rainbow-AdS$_3$ background in 3D Chern-Simons colored gravity~\cite{Gwak:2015vfb,Gwak:2015jdo,Joung:2017hsi} where the modes analogous to $\bmm_{(p,q)}$ become partially-massless through the Higgs mechanism.

\begin{center}
\begin{table}[t!]
\renewcommand{\arraystretch}{1.8}
\renewcommand{\thetable}{\arabic{table}}
\begin{tabular}{ |>{\centering\arraybackslash} m{8cm}|>{\centering\arraybackslash}m{7cm}| } 
\hline
Mode & Form of Quadratic Action\\[1ex]
\hline
$\bmm_{(p,p)}$\,,\; $\bmm_{(q,q)}$\;\;:\;\; $(p^2+q^2)$ modes & $\displaystyle \sum_n  n^2(n^2-1) f^\ast_{n}f_n\quad:\;\;$ stable  \\[2ex]
\hline
$\bms_{(p,q)}$ \;\;:\;\; $(2pq)$ modes & $\displaystyle \sum_n  n^2 \phi^\ast_{n}\phi_n\quad:\;\;$ stable  \\[2ex]
\hline
$\bmm_{(p,q)}$\;\;:\;\; $(2pq)$ modes & $\displaystyle-\sum_n  n^2(n^2-1) f^\ast_{n}f_n\quad:\;\;$ unstable  \\ [2ex]
\hline
$\bmt_{(p,p)}$, $\bmt_{(q,q)}$\;\;:\;\; $[p^2+q^2-1]$ modes & $\displaystyle-\sum_n  n^2 \phi^\ast_{n}\phi_n\quad:\;\;$ unstable  \\[2ex]
\hline
\end{tabular}
\caption{The form of quadratic actions of each boundary mode. }\label{tab:mode}
\end{table}
\end{center}

\vspace{-10mm}

%%%%%%%%%%%%%%%%%%%%%%%%%%%%%%%%%%%%%%%%%%%%%%%%%%
%%%%%%%%%%%%%%%%%%%%%%%%%%%%%%%%%%%%%%%%%%%%%%%%%%
\section{Discussion}
\label{sec: discussions}
%%%%%%%%%%%%%%%%%%%%%%%%%%%%%%%%%%%%%%%%%%%%%%%%%%
%%%%%%%%%%%%%%%%%%%%%%%%%%%%%%%%%%%%%%%%%%%%%%%%%%

In this work we have studied the boundary effective action of the colored JT gravity for the nearly-colored-AdS$_2$. Starting from the $su(N,N)$ BF theory we have derived the boundary effective action which is the color generalization of the Schwarzian theory. We have also obtained the isometry of the colored JT gravity and have shown that due to the redundant description of the boundary degrees of freedom  this isometry in the boundary effective action should be modded out.  Furthermore, we have proposed the boundary action of the colored JT gravity at finite temperature by using 
the Iwasawa-like decomposition. 

Also, we have investigated the colored asymptotic AdS$_2$ symmetry. The perturbative analysis of the boundary action have revealed  the instability of the spin-1 mode in the colored JT gravity. In particular, we have demonstrated the influence of the instability on the quantum chaos measured by the Lyapunov exponent of the OTOC. We have also proposed the rainbow-AdS$_2$ geometry where the $SU(N)$ color gauge symmetry is spontaneously broken and have obtained the boundary effective action along with  the asymptotic rainbow-AdS$_2$ symmetry.

Since the instability is originated from the spin-1 sector of the theory one can truncate it from the boundary action~\eqref{eq: boundary action zero temp} to construct the matrix generalization of the Schwarzian action without the instability:
\begin{align}
    S_{\text{\tiny mSch}}\,\equiv \, -{\kappa \gamma \over 2N } \int \rd u\;\tr\bigg[\dot {\bm f}^{-1}\,\dddot{\bm f}
    -{3\over 2}\,(\dot{\bm f}^{-1}\ddot{\bm f})^2\bigg]\,. \label{eq: matrix schwarzian action}
\end{align}
It is intriguing to ask whether the one-loop exactness of the $sl(2,\mathbb{R})$ Schwarzian action still holds in the matrix-Schwarzian action~\eqref{eq: matrix schwarzian action}. This might also be related to the question of whether the matrix-Schwarzian action can be interpreted as a coadjoint orbit action of a certain group. Such questions require the matrix-Schwarzian at finite temperature~\eqref{eq: matrix schwarzian lag at finite t}
\begin{align}
    S_{\text{\tiny mSch}}\,\equiv \, -{\kappa \gamma \over N } \int \rd u\;\tr\left[\frac{d}{du}(\bm b^{-1}\,\dot{\bm b})
	- \left(\frac{\bm b^{-1}\partial_L{\bm b}+\partial_R{\bm b}\,\bm b^{-1}}2\right)^{\!\!2}
	-\bm b^4
	\right]\,. \label{eq: matrix schwarzian action at finite temp}
\end{align}
But, as explained in Section~\bref{sec: finite temperature}, we found it difficult to express the matrix-Schwarzian at finite temperature~\eqref{eq: matrix schwarzian action at finite temp} in an explicit closed form in terms of a single variable $\bmf(u)$ unlike the zero-temperature case~\eqref{eq: matrix schwarzian action}. Nevertheless, the perturbative analysis can still be applied to the matrix-Schwarzian theory at finite temperature, and, hopefully, it can shed light on its structure. 

The instability of a quantum mechanical model can often be cured by supersymmetry. Hence, it is natural to consider a supersymmetric generalization of the colored JT gravity. Although the supersymmetry would not be helpful in avoiding the instability from the point of view of gravity, it is worthwhile to check the instability of the supersymmetric one-dimensional boundary action if exists. In addition, it would be interesting to study the higher-spin extension of the colored JT gravity as in the three-dimensional colored higher-spin gravity~\cite{Gwak:2015jdo}. In spite of the instability, those extensions will provide a fruitful testing ground for algebraic structures induced by the presence of  extended space-time symmetries.

We have shown that the color symmetry broken by the rainbow-AdS$_2$ background leads to $2pq$ unstable spin-2 modes $\bmm_{(p,q)}$ and $2pq$ stable spin-1 modes $\bms_{(p,q)}$. In $AdS_3$ such modes become partially-massless by virtue of the Higgs mechanism~\cite{Gwak:2015vfb,Gwak:2015jdo,Joung:2017hsi}. In our case, it is not clear how to define a notion of ``partially-massless'' fields because of the degenerate classification of higher-spin fields in $AdS_2$.\footnote{Moreover, all higher-spin fields in two dimensions can be interpreted in some sense as ``partially-massless fields of the maximal depth'' \cite{Alkalaev:2013fsa}.} At least, we have confirmed that the kinematic nature of the analogous modes is changed by the rainbow-AdS$_2$ background. It will be interesting to investigate further the Higgs mechanism and the resulting ``partially-massless'' modes for the colored JT gravity. Furthermore, unlike the $\bm Z =\bmI$ case, the non-singlet part of $\bmcL$ can appear in the boundary action, which is reminiscence of the chemical potential in the higher-spin gravity~\cite{Gutperle:2011kf,Ammon:2011nk,Castro:2011fm,Henneaux:2013dra}. Such a boundary effective action at finite temperature could be derived from an Iwasawa-like decomposition of the group elements which define the rainbow-AdS$_2$ background connection. 

The boundary effective action of the colored JT gravity has multiple branches which  correspond to different solutions of  the matrix equation $\dot{\bm f} =\bm b^2$ (or the analogous  equation in the finite-temperature case). In the uncolored case, there are also two branches corresponding to the different signs of $c$ but these  are isomorphic to each other and the $\mathbb Z_2$ quotient of $PSL(2,\mathbb R)=SL(2,\mathbb R)/\mathbb Z_2$ identifies them. In the colored case, different branches have different natures, in particular, different branches have different dimensions. The relevant issue  has been addressed  within the context of the colorful particles in 3D  \cite{Gomis:2021irw}, where such branches arise from the  matrix mass-shell equation $\bm P^2=m^2\ \bm I$. In the present case, the perturbative dynamics has been analyzed  for the branch connected to the solution with $\bm b=\bm I$. For other branches one would get  different dynamics, in particular, different numbers of physical degrees of freedom.

Finally, it would be interesting to identify a quantum mechanical model of which the low-energy sector is described by the boundary effective action of the colored JT gravity~\eqref{eq: boundary action zero temp}. However, such a holographic model, if exists, would also suffer from the instability. Instead, one can consider the matrix-Schwarzian action~\eqref{eq: matrix schwarzian action} as the low-energy effective action and look for the generalizations of the SYK-like models to incorporate the ``colored reparameterization symmetry'' at the infinite coupling constant limit. We leave those questions for future work.

%%%%%%%%%%%%%%%%%%%%%%%%%%%%%%%%%%%%%%%%%%%%%%%%%%
%%%%%%%%%%%%%%%%%%%%%%%%%%%%%%%%%%%%%%%%%%%%%%%%%%

\vspace{3mm}

\paragraph{Acknowledgements.} We are grateful to Xavier Bekaert, Cheng Peng, Dongwook Ghim for discussions. E.J. thanks Joaquim Gomis for his encouragement of
the current work. The work of K.A. was supported by the Foundation for the Advancement of Theoretical Physics and Mathematics “BASIS”. 
The work of E.J. was supported by National Research Foundation (Korea) through the grant NRF-2019R1F1A1044065.
The work of J.Y. was supported by KIAS individual Grant PG070102 at Korea Institute for Advanced Study and the National Research Foundation of Korea (NRF) grant funded by the Korea government (MSIT) (No. 2019R1F1A1045971, 2022R1A2C1003182). J.Y. is supported by an appointment to the JRG Program at the APCTP through the Science and Technology Promotion Fund and Lottery Fund of the Korean Government. J.Y. is also supported by the Korean Local Governments -- Gyeongsangbuk-do Province and Pohang City.
   
\appendix 

\section{Matrix relations from Section \bref{sec:rainbow}.}
\label{app:rainbow}

\noindent ${\bf 1.}$ Block-diagonal matrix notation: for a given $N\times N$ matrix $\bm M$, let $\bm M_{(a,b)}$ ($a,b=p,q$) denote a submatrix of $\bm M$ such that
\begin{align}
\label{M_block}
    \bm M\,=\, \begin{pmatrix}
    \bm  M_{(p,p)} & \bm  M_{(p,q)}\\
    \bm M_{(q,p)} & \bm M_{(q,q)}\\
    \end{pmatrix}. 
\end{align}

\noindent ${\bf 2.}$ Constant gauge transformation \eqref{const_gauge_tr}:
\begin{align} 
    \bm \cL'\,=\,&
    \bm B^\dagger \,\bm Z\,\bm B
    -i\,\bm D^\dagger\,\bm\cW\,\bm B
    +i\,\bm B^\dagger\,\bm\cW^\dag\,\bm D
    +\bm D^\dagger\,\bm\cL\,\bm D\,,\label{this_eq1}\\ 
     \bm \cW'\,=\,&
    -\bm B^\dagger \,\bm Z\,\bm A
    +i\,\bm D^\dagger\,\bm\cW\,\bm A
    -i\,\bm B^\dagger\,\bm\cW^\dag\,\bm C
    -\bm D^\dagger\,\bm\cL\,\bm C\,, \\ 
    \bm Z'\,=\,&
    \bm A^\dagger \,\bm Z\,\bm A
    -i\,\bm C^\dagger\,\bm\cW\,\bm A
    +i\,\bm A^\dagger\,\bm\cW^\dag\,\bm C
    +\bm C^\dagger\,\bm\cL\,\bm C
    \,.\label{this_eq4}
\end{align}

\noindent ${\bf 3.}$ The gauge transformation \eqref{rain_gauge_tran}:  
\be
\label{eq: transf of a rainbow}
\begin{alignedat}{2}
    &\delta a(u)\,=\, \dot{h} + [a(u),h] \\
    &\,=\, \begin{pmatrix}
    \dot{\bml} & \dot{\bmn} \\
    \dot{\bmm} & -\dot{\bml}^\dag \\
    \end{pmatrix}+ \begin{pmatrix}
    -\bmn\bmZ - \bmcL \bmm +i\, [\bm \cW,\bml ] &  \bmcL \bml^\dag + \bml\bmcL +i\left( \bm \cW\, \bmn - \bmn\, \bm \cW^\dagger\right) \\
    \bmZ \bml+\bml^\dag \bmZ+  i\left(\bm \cW^\dagger \, \bmm - \bmm\, \bm \cW\right)  & \bmZ \bmn+\bmm\bmcL - i\,[\bm \cW^\dagger,\bml^\dag]  \\
    \end{pmatrix}.
\end{alignedat}
\ee

\noindent ${\bf 4.}$ Asymptotic rainbow symmetry constraints:
\begin{align}
    \bmZ \bml&+\bml^\dag \bmZ \,=\, 2\begin{pmatrix}
    \bms_{(p,p)} & i\bmt_{(p,q)}\\
    -i\bmt_{(q,p)} & -\bms_{(q,q)}\\
    \end{pmatrix} \,=\,
    -\dot{\bmm}
    + i\left(\bmm\, \bm \cW 
    - \bm \cW^\dagger \, \bmm\right)\,,  \label{eq: asym rads sol 1}
\end{align}
\begin{align}
&\bmn_{(p,p)}\, =\, \big(\dot{\bms} - \half\{\bmcL,\bmm\}\big)_{(p,p)} 
+i [\bmcJ_{(p,p)},\bms_{(p,p)}] 
+ i \big(\bm \cJ_{(p,q)} \, \bmt_{(q,p)}- \bmt_{(p,q)}\,\bm \cJ_{(q,p)}\big)\ ,\\
&\bmn_{(q,q)}\,=\,\big(-\dot{\bms} + \half\{\bmcL,\bmm\}\big)_{(q,q)} - i\big[\bmcJ_{(q,q)}, \bms_{(q,q)}\big] 
- i \big(\bm \cJ_{(q,p)} \, \bmt_{(p,q)}-\bmt_{(q,p)}\,\bm \cJ_{(p,q)}\big)\ ,\\
&\begin{alignedat}{2}
\bmn_{(p,q)}\,=\,&\big(-i\dot{\bmt} + \half \big[\bmcL, \bmm\big]\big)_{(p,q)} + \big(\bm \cJ_{(p,p)} \, \bmt_{(p,q)} - \bmt_{(p,q)}\,\bm \cJ_{(q,q)}\big)
\vspace{2mm}
\\
&- \big(\bm \cJ_{(p,q)} \, \bms_{(q,q)}- \bm  \bms_{(p,p)}\,\bm \cJ_{(p,q)}\big)\,,
\end{alignedat}\\
&\begin{alignedat}{2}
\bmn_{(q,p)}\,=\,&\big(i\dot{\bmt} - \half[\bmcL,\bmm]\big)_{(q,p)} - \big(\bm \cJ_{(q,q)} \, \bmt_{(q,p)} - \bm  \bmt_{(q,p)}\,\bm \cJ_{(p,p)}\big)
\vspace{2mm}
\\
&+  \big(\bm \cJ_{(q,p)} \, \bms_{(p,p)}- \bm  \bms_{(q,q)}\,\bm \cJ_{(q,p)}\big)\ .\label{eq: asym rads sol 5}
\end{alignedat}
\end{align}

\noindent {\bf 5.}
Let us now calculate each term in the quadratic action \eqref{eq: bdy action rainbow2}. At order $\cO(\epsilon^1)$, we find 
\be
\label{L1}
\begin{alignedat}{2}
\bmcL^{(1)}\,=\,&-\dot{\bmn} -{\pi^2\over \beta^2} (\bml \bmZ + \bmZ \bml^\ast)
\,=\,
\begin{pmatrix}
	\half \ddot{\bmm}_{(p,p)} + {2\pi^2\over \beta^2} \dot{\bmm}_{(p,p)}& -\half\ddot{\bmm}_{(p,q)} - {2\pi^2\over \beta^2} \dot{\bmm}_{(p,q)} \\
	-\half  \ddot{\bmm}_{(q,p)} - {2\pi^2\over \beta^2} \dot{\bmm}_{(q,p)}& \half \ddot{\bmm}_{(q,q)}+ {2\pi^2\over \beta^2} \dot{\bmm}_{(q,q)}
	\end{pmatrix},
\vspace{3mm}
\\
    \bm\cW^{(1)}\,=\,& \begin{pmatrix}
	\dot{\bmt}_{(p,p)} & -i\dot{\bms}_{(p,q)}\\
	-i\dot{\bms}_{(q,p)}& \dot{\bmt}_{(q,q)}\\
	\end{pmatrix}.
\end{alignedat}
\ee
At order $\cO(\epsilon^2)$, we will also demand the asymptotic rainbow-AdS condition~\eqref{eq: asymptotic rainbow ads condition} together with the gauge condition~\eqref{eq: colored gauge field a} to determine $k$ in \eqref{gauge_p_hk}. Especially, by demanding $\bmZ$ part of $a$ in \eqref{eq: general form of gauge field a}, the higher-order gauge parameters $\bm \sigma_{(p,p)}, \bm \sigma_{(q,q)}, \bm \tau_{(p,q)}$ and $\bm\tau_{(q,p)}$ in \eqref{gauge_p_hk} can be obtained as follows:
\be
\begin{alignedat}{2}
    &\begin{pmatrix}
	\bm \sigma_{(p,p)} & i\bm \tau_{(p,q)}\\
	-i\bm \tau_{(q,p)} & - \bm \sigma_{(q,q)}\\
	\end{pmatrix}
	\,=\,- \half \begin{pmatrix}
	\dot{\bm \mu}_{(p,p)} & \dot{\bm \mu}_{(p,q)}\\
	\dot{\bm \mu}_{(q,p)} & \dot{\bm \mu}_{(q,q)}\\
	\end{pmatrix}- \half\left(\dot{\bmm}\bml - \dot{\bml}^\dag \bmm\right) + \half\left(\bmm\dot{\bml} - \bml^\dag \dot{\bmm}\right)
\vspace{3mm}
\\
 &\hspace{30mm}-\half\left[ \bmZ( \bml^2+\bmn \bmm) +  \left((\bml^\dag)^2+ \bmm \bmn\right)\bmZ \right]- \left(\bml^\dag \bmZ \bml +{\pi^2\over \beta^2} \bmm\bmZ \bmm \right),\label{eq: rainbow ads 2nd order sol}
\end{alignedat}
\ee
where $\bml\equiv \bms+i \bmt$. Similarly, $\bmcL^{(2)}$ is found to be
\be
\begin{alignedat}{2}
    &\bmcL^{(2)}\,=\,  -{1\over 2} \dot{\bm\nu}-{\pi^2\over 2\beta^2}\big( \{\bm \sigma, \bmZ\} + i [ \bm \tau, \bmZ] \big)\\
    &+{1\over 2} \big( \{\bm s, \dot{\bm n} \} - \{\dot{\bm s}, \bm n\} \big)+ {i\over 2} \big( [\bm t, \dot{\bm n} ] - [\dot{\bm t}, \bm n] \big) + \bm n \bmZ \bm n\\
    &+ {\pi^2\over 2\beta^2} \bigg( \bmZ \bm m \bm n + \bm n \bm m \bmZ  
    +\bm l^2\,\bm Z+\bm Z\,\bm l^\dagger{}^2
     +2 \bm l \bmZ \bm l^\dagger  \bigg)\ .
\end{alignedat}
\ee
Note that one could also determine $\bm \nu$ from  $\bm\cW$ and $\bm\cW^\dag$. However, the solution for $\bm \nu$ is not necessary in evaluating the quadratic action because the term related to $\bm\nu$ in $\bmcL^{(2)}$ is total derivative. Using \eqref{eq: rainbow ads 2nd order sol} as well as \eqref{eq: asym rads sol 1}--\eqref{eq: asym rads sol 5} we find the respective part of the effective action  $\tr [\bmZ \bmcL^{(2)}]$ \eqref{eq: bdy action rainbow} up to total derivatives:
\be
\label{L2}
\begin{alignedat}{2}
	\int \rd u\;\tr [\bmZ \bmcL^{(2)}]
	\,=\, &  \int \rd u\; \tr\bigg[ {\pi^2\over \beta^2}\, \bmZ( \dot{\bmm}\bml+\bml^\dag \dot{\bmm}) - \bmZ(\dot{\bml} \bmn+ \bmn \dot{\bml}^\dag)\bigg] \\
	&+\int \rd  u\; \tr\bigg[ {\pi^2\over \beta^2}\,\bigg( \bmZ \bml^\dag\bmZ \bml +{\pi^2\over \beta^2} (\bmZ \bmm)^2\bigg) + (\bmZ \bmn)^2 +{\pi^2\over \beta^2} \bmZ \bml \bmZ \bml^\dag \bigg]\\
	&+ {2\pi^2\over \beta^2} \int \rd  u\; \tr \big[\bms^2 - \bmt^2 + \bmm \bmn  \big]\,. 
\end{alignedat}
\ee

\bibliographystyle{JHEP}
\bibliography{color_schwarzian}

\end{document}